\documentclass{article}
\usepackage[utf8]{inputenc}
\usepackage{amsmath}
\usepackage{booktabs}
\usepackage{multirow}
\usepackage{float}
\usepackage{bm}
\usepackage{graphicx}
\makeatletter

\newcommand{\Rmnum}[1]{\expandafter\@slowromancap\romannumeral #1@}
\makeatother
\usepackage{color}
\usepackage{algorithm}
\usepackage{algorithmic}
\usepackage{amssymb}
\usepackage{subfigure}
\usepackage{caption}
\usepackage{authblk}
\graphicspath{{fig/Time/}}
\usepackage{hyperref}
\usepackage{geometry}

\usepackage{natbib}

\definecolor{red}{RGB}{139,0,18}

\hypersetup{hypertex=true,
    colorlinks=true,
    linkcolor=red,
    anchorcolor=green,
    citecolor=blue}

\title{\textbf{Unified Distributed Estimation Framework for Sufficient Dimension Reduction Based on Conditional Moments}}
\author[a]{Hongying Li}
\author[b]{Minyi Zhu}
\author[c]{Yaqi Cao}
\author[b,*]{Xinyi Xu}
\affil[a]{\small Department of Statistics, The Ohio State University, Columbus, Ohio, United States of America}
\affil[b]{\small School of Statistics and Mathematics, Central University of Finance and Economics, Beijing, China}
\affil[c]{\small School of Science, Minzu University of China, Beijing, China}
\affil[*]{\small Corresponding Author: xinyixu@cufe.edu.cn}
\date{}

\begin{document}
\maketitle

\begin{abstract}
Nowadays, massive datasets are typically dispersed across multiple locations, encountering dual challenges of high dimensionality and huge sample size. Therefore, it is necessary to explore sufficient dimension reduction (SDR) methods for distributed data. In this paper, we first propose an exact distributed estimation of sliced inverse regression, which substantially improves computational efficiency while obtaining identical estimation as that on the full sample. Then, we propose a unified distributed framework for general conditional-moment-based inverse regression methods. This framework allows for distinct population structure for data distributed at different locations, thus addressing the issue of heterogeneity. To assess the effectiveness of our proposed methods, we conduct simulations incorporating various data generation mechanisms, and examine scenarios where samples are homogeneous equally, heterogeneous equally, and heterogeneous unequally scattered across local nodes. Our findings highlight the versatility and applicability of the unified framework. Meanwhile, the communication cost is practically acceptable and the computation cost is greatly reduced. Sensitivity analysis verifies the robustness of the algorithm under extreme conditions where the SDR method locally fails on some nodes. A real data analysis also demonstrates the superior performance of the algorithm.
\end{abstract} 


\section{Introduction}
Sufficient dimension reduction (SDR) aims to replace the high-dimensional explanatory variables with a small number of linear projections that preserve all information about the response variable. This property can be expressed by 
\begin{equation}\label{eq1}
    \ Y \rotatebox[origin=c]{90}{$\vDash$} \bm{X}\mid \bm{\beta}^T \bm{X},\
\end{equation}
where $\bm{X}\in\mathbb{R}^p$ is the predictor vector, $Y\in\mathbb{R}$ is the univariate response, $\bm{\beta}$ is a $p \times k$ matrix with $k \leq p$, and $\rotatebox[origin=c]{90}{$\vDash$}$ stands for independence. The column subspace of $\bm{\beta}$ with minimum dimension is often called the central subspace (CS, \citet{cook2009regression}) and denoted by $\mathcal{S}_{Y|X}$. The dimension of $\mathcal{S}_{Y|X}$, indicated by $d$, is called the structural dimension, which is usually unknown and far less than $p$ in high-dimensional scenarios. In this way, the dimension of the predictor can be reduced to avoid the ``curse of dimensionality'' and meanwhile improve computational efficiency.
Classical sufficient dimension reduction methods are divided into inverse regression, forward regression, and semi-parametric regression. Inverse regression $\bm{X}|Y$ generally has two paradigms. One is the moment-based approach, and the other is the model-based approach. Moment-based reduction is model-free and uses conditional moments to estimate the central subspace. Among them, sliced inverse regression (SIR, \citet{li1991sliced}) was first proposed by using the inverse mean $E(\bm{X}|Y)$. Subsequent approaches including sliced average variance estimation (SAVE, \citet{cook1991sliced}), the conditional $k$th moment method \citep{yin2002dimension}, the inverse third moment method \citep{yin2003estimating}, contour regression (CR, \citet{li2005contour}) and directional regression (DR, \citet{li2007directional}) are based on a higher-order conditional moment.

With the development of information technology, most massive data is now stored in multiple locations due to issues such as security, privacy, and the limited storage capacity of individual servers. This distributed system poses dual challenges of high dimensionality and huge sample size. To address the \textbf{dimension reduction} problems of massive distributed data, some literature focuses on identifying the explanatory variables that truly impact the response variable $Y$ based on the assumption of sparsity, using either parametric or linear models, as described in \citet{chen2014split}, \citet{lee2017communication}, \citet{battey2018distributed}, \citet{jordan2018communication}and  \citet{fan2023communication}. In particular, \citet{li2020distributed} proposes a distributed feature screening framework to select a small subset of $\bm{X}$ that is important for predicting $Y$. Another category of literature proposes distributed algorithms for the most common unsupervised dimension reduction method: principal component analysis (PCA), including \citet{kargupta2001distributed}, \citet{qi2004global}, \citet{bai2005principal}, \citet{liang2014improved} and \citet{zhang2018dimension} . The theoretical properties of distributed algorithms for PCA have been thoroughly studied by \citet{fan2019distributed}. 

There are also several studies that focus on \textbf{sufficient dimension reduction} methods for distributed data. For example, \citet{jin2019distributed} introduces the distributed estimation of principal support vector machines (D-PSVM). \citet{zhu2022distributed} considers sufficient dimension reduction for heterogeneous massive data within a unified least squares framework. \citet{liquet2016big} proposes a new estimator called BIG-SIR, which follows a ``divide and conquer'' strategy. In this approach, SDR directions are estimated in each subset, and then a recombination step is performed based on the optimization of a criterion that assesses the proximity between the SDR directions of each subset. The consistency and the asymptotic distribution of the BIG-SIR estimator are provided, which establishes the theoretical foundation for our study. Furthermore, \citet{xu2022distributed} presents distributed sufficient dimension reduction methods for heterogeneous massive data. They explicitly examine the sliced inverse regression (SIR) and cumulative slicing estimation (CUME, \citep{zhu2010dimension}), and investigate the nonasymptotic error bounds of the SDR results.

Ideas underlying \citet{liquet2016big} and \citet{xu2022distributed} both involve aggregating local results from individual nodes to obtain a final estimate on a central server, which also inspires our research in this article. However, unlike \citet{liquet2016big}, which only revisits SIR, and \citet{xu2022distributed}, which only examines SIR and CUME, our article establishes a unified framework for distributed estimation of moment-based sufficient dimension reduction methods. Specifically, we first propose an exact distributed estimation of sliced inverse regression (E-DSIR), where simple statistics such as sample means, maximum and minimum response values are transmitted between the central and local servers through a two-round communication. This algorithm does not involve any additional approximation, ensuring exactly the same estimate as the original SIR. Then we extend the distributed SIR algorithm to general moment-based inverse regression methods, represented by SIR, SAVE, and DR. This algorithm utilizes a one-shot communication to aggregate SDR directions obtained from each local node and yields a final inverse conditional moment on the central server. Moreover, our article breaks the limitation in \citet{liquet2016big} where samples are assumed to be homogeneous uniformly scattered on local nodes. We allow for distinctive population structure for data scattered at different locations, thus addressing the issue of heterogeneity. Note that this is different from the heterogeneity in \citet{xu2022distributed}, which focuses on different function maps for data scattered at different locations. Additionally, we investigate various details of the algorithm, including the selection of local and global parameters, the relationship between local optimality and global optimality, the transmission cost and the stability in extreme scenarios. Simulation results demonstrate that the proposed distributed algorithm achieves nearly the same estimation efficiency as global SDR methods and meanwhile maintain robustness under various data generation mechanisms and partition schemes across local nodes. An application to a critical temperature dataset also verifies the effectiveness of our algorithm.

The rest of the article is organized as follows. Section \ref{sec method} revisits classical conditional-moment-based SDR methods and proposes two distributed estimations. Section \ref{sec Simulation} and Section \ref{sec real} examine the performance of our proposed algorithms through a wide range of simulations and real data analysis. Section \ref{sec conclude} concludes the paper with a brief discussion.

\section{Methods and Algorithms}\label{sec method}

\subsection{Preliminaries}
\subsubsection{Conditional-Moment-Based Inverse Regression}

In this section, we first review classical moment-based inverse regression methods for sufficient dimension reduction, represented by SIR, SAVE and DR. 

To recover $\mathcal{S}_{Y|X}$, \citet{li1991sliced} proposed sliced inverse regression (SIR) by using the inverse mean $E(\bm{X}|Y)$ through the slicing estimation procedure. The slicing estimation is a simple and computationally efficient nonparametric estimation. SIR targets $\mathcal{S}_{Y|X}$ through spectral decomposition on the matrix $\bm{\Sigma}^{-1}\bm{V}$, where $\bm{\Sigma}=:var(\bm{X})$ can be estimated by the sample variance matrix and $\bm{V}=:var\{E(\bm{X}|Y)\}$. SIR divides the range of $Y$ into several slices, then computes the sample mean of the predictors within each slice. To be precise, given observations $\{\bm{x}_i, y_i\}$, $i=1,...,n$, SIR (usually equally) partitions the range of $Y$ into $H$ slices $\{I_1,\cdots,I_H\}$, where $I_h$ denotes the interval $(\tilde{y}_{h-1},\tilde{y}_h]$ and $\tilde{y}_0<\tilde{y}_1<\cdots\tilde{y}_H$ are user-specified grid points. The weighted variance matrix
\begin{equation}\label{eq2.1}
    \hat{\bm{V}}=\sum_{h=1}^H E\{\bm{X}\bm{1}(Y\in I_h)\}E\{\bm{X}^T\bm{1}(Y\in I_h)\}/E\{\bm{1}(Y\in I_h)\}
\end{equation}
provides a consistent estimation of $var\{E(\bm{X}|Y)\}$, where $\bm{1}(\cdot)$ is an indicator function. We pay particular attention to SIR because it is perhaps the most widely used method for estimating $\mathcal{S}_{Y|X}$ due to its computational simplicity. 

However, SIR sometimes fails when $E(\bm{X}|Y)$ degenerates. \citet{cook1991sliced} and \citet{li2007directional} developed sliced average variance estimation (SAVE) and directional regression (DR) respectively by using additionally the inverse variance $var(\bm{X}|Y)$, which applied the slicing estimation based on a higher-order conditional moment. SAVE recovers $\mathcal{S}_{Y|X}$ through the column space of $\bm{\Sigma}^{-1}\bm{V}$, where $\bm{V}=:E\{var(\bm{X})-var(\bm{X}|Y)\}^2$, which is approximated by
\begin{equation}\label{eq2.2}
    \hat{\bm{V}}=\sum_{h=1}^H\{var(\bm{X})-var(\bm{X}|Y\in I_h)\}^2E\{\bm{1}(Y\in I_h)\}.
\end{equation}
DR provides information about $\mathcal{S}_{Y|X}$ by $\bm{V}=:E\{(\Tilde{\bm{X}}-\bm{X})(\Tilde{\bm{X}}-\bm{X})^T|\Tilde{Y},Y\}$, where $(\Tilde{\bm{X}},\Tilde{Y})$ is an independent copy of $(\bm{X},Y)$, which is approximated by
\begin{equation}\label{eq2.3}
\begin{aligned}
    \hat{\bm{V}}=&2\sum_{h=1}^H[E(\bm{ZZ}^T|Y\in I_h)]^2E[\bm{1}(Y\in I_h)]\\
    +&2\left\{\sum_{h=1}^H[E(\bm{Z}|Y\in I_h)E(\bm{Z}^T|Y\in I_h)]E[\bm{1}(Y\in I_h)]\right\}^2\\
    +&2\left\{\sum_{h=1}^H[E(\bm{Z}^T|Y\in I_h)E(\bm{Z}|Y\in I_h)]E[\bm{1}(Y\in I_h)]\right\}\\
    &\cdot\left\{\sum_{h=1}^H[E(\bm{Z}|Y\in I_h)E(\bm{Z}^T|Y\in I_h)]E[\bm{1}(Y\in I_h)]\right\}\\
    -&2\bm{I}_p,
\end{aligned}
\end{equation}
where $\bm{Z}$ is the standardized predictor.

Given different forms of the conditional moment with its slicing estimation, it is commonly advocated to extract the first $K$ eigenvectors (in order of eigenvalues) as a favorable estimator of sufficient dimension reduction directions, where $K$ is supposed to be $\geq d$. 

However, classical slicing estimation of the inverse moment can be computationally difficult and very time-consuming when generalized to massive datasets. Moreover, the number of slices should grow with the increase of sample size if we want to construct the estimation consistency of the target matrices, which further increases the computation burden. Therefore, we tend to scatter the whole sample into several servers and reconstruct these conditional-moment-based SDR methods in the distributed system.

\subsubsection{Sliced Inverse Regression for Massive Data}
\citet{liquet2016big} proposes BIG-SIR estimator based on the divide-and-conquer (DAC) principle to deal with massive datasets, which consists in: (i) divide the massive dataset into $G$ blocks; (ii) apply the usual SIR estimator on each block separately; (iii) recombine the SDR directions from each block to get a solution of the full data. The recombination step is based on the optimization of a criterion which assesses the proximity between the SDR directions of each subset. The optimization problem is defined as
\begin{equation}\label{eq5}
    \max_{\bm{A}} \sum_{g=1}^{G} w_g m(\bm{A},\bm{B}_g) \quad \text{s.t.} \quad \bm{A^T} \bm{A}=\bm{I}_K.
\end{equation}
$\bm{A}$ is a $p\times K$ matrix. $\bm{B}_1$, $\bm{B}_2$, \ldots, $\bm{B}_G$ are defined as the $\bm{I}_p$-orthonormal basis of the SDR space obtained from each block. $w_1$, $w_2$, \ldots, $w_G$ are positive weights such that $\sum_{g=1}^{G} w_g = 1$, which allows the algorithm to take into account different block sizes. $m(\bm{A}, \bm{B}_g)$ measures the proximity between two linear subspaces spanned by $\bm{A}$ and $\bm{B}_g$, which can be defined by their trace correlation. The trace correlation is specifically defined in Section \ref{sec evalue metric} and takes its values in [0,1].

\hspace*{\fill}

$\textit{\textbf{Theorem.}}$ \textit{The solution $\bm{B}$ of the maximization problem (\ref{eq5}) is an $\bm{I}_p$-orthonormal basis of the K-dimensional eigenspace associated with the K largest eigenvalues $\lambda_1$, \ldots, $\lambda_k$ of the $p\times p$ matrix
$$\bm{V} = \sum_{g=1}^{G} w_g \left(\frac{\bm{B}_g \bm{B}_g^T}{K}\right).$$
Moreover, $\sum_{g=1}^{G} w_g m(\bm{B}, \bm{B}_g) = \lambda_1 + \ldots + \lambda_K.$ Under the linearity condition and assumption (\ref{eq1}), the column vectors of $\bm{B}$ form an $\bm{I}_p$-orthonormal basis of $\mathcal{S}_{Y|X}$.}

\hspace*{\fill} 

\citet{liquet2016big} estimates $\hat{\bm{B}}$ by substituting the moment $\bm{B}_g$ with its sample version $\hat{\bm{B}_g}$, and therefore to obtain the estimation of the SDR subspace.

The linearity condition requires that $E(\bm{X}\mid \bm{B}^T \bm{X})$ is a linear function of $\bm{X}$, which is satisfied when $\bm{X}$ follows a normal or, more generally, an elliptically contoured distribution. \citet{xu2022distributed} regards the linearity condition as mild since this condition holds asymptotically, as long as $p$ is sufficiently large and the intrinsic dimension $d$ is relatively small. More importantly, \citet{xu2022distributed} investigates the nonasymptotic error bounds for the distributed estimation $\hat{\bm{B}}$ and concludes that the error bound of $\hat{\bm{B}}$ is minimized when the number of local nodes is of order $(n/p)$. In this paper, we extend their distributed SIR theory to general moment-based inverse regression. Please see Section \ref{sec algorithm2} for detail.

\subsection{Distributed Sufficient Dimension Reduction}

\subsubsection{Exact Distributed  Estimation of Sliced Inverse Regression}

The original algorithm of SIR is shown in Algorithm~\ref{alg:original SIR} in Appendix. 
An eigen-decomposition is conducted on the weighted covariance matrix $\hat{\bm{V}}$ to track the first $K$ eigenvectors as the SDR directions. 
However, in the distributed system, Algorithm~\ref{alg:original SIR} can not be directly applied to estimate the covariance matrix $\bm{V}$, as the data are distributed among several distant machines (called ``workers''). To solve this problem, we design the exact distributed estimation of sliced inverse regression (E-DSIR) in Algorithm~\ref{alg:dis SIR} to intelligently calculate the weighted covariance matrix. 

Two rounds of worker-master communications are involved in the algorithm. In the first round, we compute the sample mean of $\bm{x}$ and the range of $y$ within each worker, summarize these statistics, and obtain the global sample mean of $\bm{x}$ and the total range of $y$ on the master node. The total range of $y$ enables us to determine the range of $H$ slices, which is transmitted back to each worker. In the second round, we calculate the sample mean and size within each slice on each worker. The results are sent to the master node to gain the sample mean and size within each slice in the global scale. With the required values available, we generate the weighted covariance matrix $\hat{\bm{V}}$ and conduct eigen-decomposition on it. The weighted covariance matrix $\hat{\bm{V}}$ obtained in this algorithm is exactly the same as that in Algorithm~\ref{alg:original SIR}, thus yielding identical estimates of SDR directions $\hat{\bm{\beta}}_k$ ($k=1,...,K$) as the original SIR.

\begin{algorithm}[H] 
\caption{Exact Distributed Estimation of Sliced Inverse Regression} 
\label{alg:dis SIR} 
\begin{algorithmic}[1] %1 means every line will show row number.
\REQUIRE ~~\\ %Input
Observations $\{(y_i, \bm{x}_i), i=1,...,n\}$ scattered on S workers: $W_1, ..., W_S$.
\ENSURE ~~\\ %Output
\STATE For $s = 1, ..., S$, within each worker, obtain the minimum and maximum values of $\{y_i, i\in W_s\}$, marked as $y_{s_\text{min}}$ and $y_{s_\text{max}}$. Calculate the sample mean $\bar{\bm{x}}_s=\sum\limits_{i\in W_s}\bm{x}_i/n_s$, where $n_s$ is the sample size on worker $s$. Transmit statistics $y_{s_\text{min}}$, $y_{s_\text{max}}$ and $\bar{\bm{x}}_s$ to the master node.
\STATE On the master, obtain the minimum and maximum values of $\{y_i, i=1,...,n\}$ based on $\{(y_{s_\text{min}}, y_{s_\text{max}}), s=1,...,S\}$, marked as $y_\text{min}$ and $y_\text{max}$. Set $H$ slices $\{I_1,\cdots,I_H\}$, where $I_h$ denotes the interval $(\tilde{y}_{h-1},\tilde{y}_h]$ and $y_\text{min}=\tilde{y}_0<\tilde{y}_1<...<\tilde{y}_H=y_\text{max}$ are grid points.
\STATE Calculate the global sample mean by $\bar{\bm{x}}=\sum_{s=1}^S\frac{n_s}{n}\bar{\bm{x}}_s$. Transmit the grid points and $\bar{\bm{x}}$ back to each worker.
\STATE For $s = 1, ..., S$, within each worker, centralize $\{\bm{x}_i, i\in W_s\}$ by computing $\bm{u}_i = \bm{x}_i - \bar{\bm{x}}$, and calculate the sample variance matrix as $\hat{\bm{\Sigma}}_s=\frac{1}{n_s-1}\sum\limits_{i\in W_s}\bm{u}_i\bm{u}_i^T$. On worker $s$, for $h=1,...,H$, calculate the sample mean within each slice as $m_{s_h}=\sum\limits_{i\in W_s,y_i\in I_h} u_i/n_{s_h}$, where $n_{s_h}$ is the sample size on worker $s$ within slice $h$. Transmit statistics $m_{s_h}$ and $\hat{\bm{\Sigma}}_s$ to the master node.
\STATE On the master, for $h=1,...,H$, calculate the sample mean within each slice as $\hat{\bm{m}}_h = \sum_{s=1}^S\frac{n_s}{n}m_{s_h}$. Denote the proportion of observations falling in slice $h$ as $\hat{p}_h = (\sum_{s=1}^S n_{s_h})/n$. Calculate the global sample variance matrix as $\hat{\bm{\Sigma}}=\frac{1}{n-1}\sum_{s=1}^S (n_s-1)\hat{\bm{\Sigma}}_s$.
\STATE Generate the covariance matrix $\hat{\bm{V}} = \sum_{h=1}^H \hat{p}_h \hat{\bm{m}}_h\hat{\bm{m}}_h^T$. Obtain the eigenvalues and eigenvectors for $\hat{\bm{V}}$. Eigenvectors corresponding to the top $K$ largest eigenvalues are denoted as $\hat{\bm{\eta}}_k$ ($k = 1, ..., K$). 
\STATE \textbf{Return} estimates of SDR directions as $\hat{\bm{\beta}}_k=\hat{\bm{\Sigma}}^{-1}\hat{\bm{\eta}}_k$ ($k = 1, ..., K$).
\end{algorithmic}
\end{algorithm}

\textit{\textbf{Communication Cost.}} If no local computation is done, the original Algorithm~\ref{alg:original SIR} that begins by centralizing all the data requires $O(np)$ data transfer. Now we discuss the communication cost of E-DSIR in two rounds. For the first round, the workers compute the local sufficient statistics, including $y_{s_\text{min}}$, $y_{s_\text{max}}$ and $\bar{\bm{x}}_s$ for $s=1,...,S$, and transfer them to the master, which requires $O(S(p+2))$ data transfer. For the second round, instead of transmitting a large $n_s\times p$ matrix, each worker sends mean vectors of $H$ slices to the master node, which is only an $H\times p$ matrix. Even though $\hat{\bm{\Sigma}}_s$ requires $O(p^2)$ data transfer, its cost is less than $O(n_sp)$ for $p<n_s$. If on some workers $n_s<p$, a simple alternative is to replace $\hat{\bm{\Sigma}}_s$ with its first $k_s$ eigenvectors, which is a fast approximate solution with $k_s\ll p$. Therefore, the communication cost in this algorithm is trivial compared with the enormous data.   

\textit{\textbf{Estimation Accuracy.}} Since no extra approximation is involved in E-DSIR compared to the original SIR (if there is no approximation of $\hat{\bm{\Sigma}}_s$), it replicates SIR into the distributed system perfectly, which ensures identical computational accuracy as global SIR. Furthermore, the algorithm does not make any assumption about the allocation of sample size or sample range at each worker node, so it is applicable to various sample distribution scenarios. In addition, the simulation verifies that the proposed algorithm can greatly improve computation efficiency (details in Section \ref{sec Computational Time}).

\subsubsection{Distributed Estimation of General Moment-Based Inverse Regression}\label{sec algorithm2}

Although E-DSIR replicates the estimate of SIR exactly to the distributed system, it is not applicable to other SDR methods. We propose a more generalized distributed estimation of the moment-based inverse regression in Algorithm~\ref{alg:dis SDR}. This paper applies this algorithm to SIR, SAVE and DR as representatives.

SIR, SAVE and DR share the similarity that they are based on the first or second conditional moments $E(\bm{X}|Y)$ and $E(\bm{X}\bm{X}^T|Y)$. Under the assumption that the inverse regression curve falls in the SDR subspace, eigen-decomposition is conducted on the inverse conditional moments to locate the SDR directions. The column space spanned by the eigenvectors is regarded as the SDR subspace. 

Inspired by the divide-and-conquer strategy, in the distributed system, we first compute the local weighted conditional matrix $\hat{\bm{V}}_s$ ($s=1,...,S$) on each worker and then conduct eigen-decomposition. For worker $s$, we only transmit $K_s$ local eigenvectors corresponding to the largest $K_s$ eigenvalues to the master node, to reduce the amount of data transfer. On the master, we reconstruct the approximation $\tilde{\bm{V}}_s$ on each worker using its corresponding eigenvectors and eigenvalues. In order to address the bias of unbalanced sample size, we take the weighted sum of these $\tilde{\bm{V}}_s$ according to the sample proportion in each worker node and generate $\tilde{\bm{V}}$. The estimates of SDR directions $\hat{\bm{\beta}}_k$ $(k = 1, ..., K_g)$ can be obtained through eigen-decomposition conducted on the global weighted conditional matrix $\tilde{\bm{V}}$.

\begin{algorithm}[H] 
\caption{Approximate Distributed Estimation of General Moment-Based Inverse Regression} 
\label{alg:dis SDR} 
\begin{algorithmic}[1] %1 means every line will show row number.
\REQUIRE ~~\\ %Input
Observations $\{(y_i, \bm{x}_i), i=1,...,n\}$ scattered on S workers: $W_1, ..., W_S$.
\ENSURE ~~\\ %Output
\STATE For $s=1,...,S$, on each worker, compute the weighted conditional matrix $\hat{\bm{V}}_s$ according to Equation (\ref{eq2.1}), (\ref{eq2.2}) and (\ref{eq2.3}) respectively for SIR, SAVE and DR.
\STATE Conduct eigen-decomposition on $\hat{\bm{V}}_s$. Specify $K_s$ as the number of principal components selected from $\hat{\bm{V}}_s$. Denote statistics $\{\lambda_{s,1},...,\lambda_{s,K_s}\}$ as the $K_s$ largest eigenvalues in descending order, and $\tilde{\bm{U}}_s=(\bm{\eta}_{s,1},...,\bm{\eta}_{s,K_s})$ as the matrix whose columns are the $K_s$ eigenvectors corresponding to the $K_s$ largest eigenvalues.
\STATE Transmit statistics $\{\lambda_{s,1},...,\lambda_{s,K_s}\}$ and $\tilde{\bm{U}}_s$ to the master node.
\STATE On the master, compute $\tilde{\bm{V}}_s=\tilde{\bm{U}}_s\bm{\Lambda}_s\tilde{\bm{U}}_s^T$ as the optimal linear approximation of $\hat{\bm{V}}_s$, where $\bm{\Lambda}_s=\text{diag}(\lambda_{s,1},...,\lambda_{s,K_s})$ is the diagonal matrix of $K_s$ largest eigenvalues.
\STATE Compute $\tilde{\bm{V}}=\sum_{s=1}^S \frac{n_s}{n} \tilde{\bm{V}}_s$ as the approximation of the global weighted conditional matrix $\hat{\bm{V}}$. Conduct eigen-decomposition on $\tilde{\bm{V}}$. 
\STATE Specify $K_g$ as the number of principal components selected from the global $\tilde{\bm{V}}$. The eigenvectors corresponding to the top $K_g$ largest eigenvalues of $\tilde{\bm{V}}$ are denoted as $\hat{\bm{\beta}}_k$ $(k = 1, ..., K_g)$.
\STATE \textbf{Return} estimates of SDR directions as $\hat{\bm{\beta}}_k$ $(k = 1, ..., K_g)$.
\end{algorithmic}
\end{algorithm}

\textit{\textbf{Choice of Dimension K.}} The local and global approximation is guided by the parameter $K_s$ and $K_g$. There are two options to set these parameters. To guarantee the quality of approximation, we can require that the proportion of total variation in the original $p$ variables explained by the $K_s$ (on worker $s$) or $K_g$ (on the master) principal components is at least $\alpha$. That is, set $(\sum_{k=1}^{K_s}\lambda_{s,k})/(\sum_{k=1}^p\lambda_{s,k})\geq\alpha$ locally and $(\sum_{k=1}^{K_g}\lambda_k)/(\sum_{k=1}^p\lambda_k)\geq\alpha$ globally, so as to determine $K_s$ and $K_g$. In this way, $K_s$ may vary on different workers for a fixed quality of approximation. To control the communication cost, we can either set $K_s$ to a fixed number on all workers, that is, fix $K_s=K$ and ensure a $p\times K$ matrix transfer. In this way, the proportion of variation explained may vary on different workers for a fixed transmission bandwidth.

\textit{\textbf{Approximation Accuracy.}} Because of linear optimality properties of principal components, the proposed algorithm combines optimal linear approximations of local weighted conditional matrices $\hat{\bm{V}}_s$. As local optimality is not sufficient for global optimality, we discuss the numerical effectiveness of the algorithm in two situations. First, if data is partitioned randomly among locations, which means samples on different workers are \textbf{homogeneous}, then global optimal results would be close to local optimal results. Particularly in Algorithm~\ref{alg:dis SDR}, different locations share a similar population with similar data scales and slice ranges, so the eigenvectors for the weighed sum $\tilde{\bm{V}}$ with respect to $\hat{\bm{\Sigma}}$ would be desirable estimates of SDR directions, which is confirmed in the following simulation and real data analysis. Second, if data is partitioned unevenly among locations, which means samples on different workers are \textbf{heterogeneous}, then each location describes a very different sub-population with a distinct sample mean and data scale. In this situation, another round of data transfer is required to transmit global $\bar{\bm{x}}$ and ranges of $H$ slices to each worker, just as steps 1-3 in Algorithm~\ref{alg:dis SIR}, where the extra communication cost is trivial. Then Algorithm~\ref{alg:dis SDR} can be performed on a globally uniform centralization and slice scale, rather than a separate centralization and slicing. 
\textbf{Despite that data among different locations has widely different scales, this heterogeneity has been thoroughly characterized by the varying range of global slices.} \citet{xu2022distributed} also demonstrates the advantage of distributed estimates over global estimates in the presence of heterogeneity, particularly when $K \geq 2$.

\textit{\textbf{Communication Cost.}} Only one shot of transmission is involved in this algorithm. Instead of sending the complete weighted conditional matrix with dimension $p\times p$, we send $K_s$ eigenvectors from worker $s$ to the master node. The total communication cost is $O(\sum_{s=1}^SK_s p)$, which is affordable because a low intrinsic dimensionality exists in most high-dimensional data settings, implying $K_s \ll p$. To further demonstrate the computation efficiency of Algorithm~\ref{alg:dis SDR}, we compare the running time of the original and distributed SDR methods in Section~\ref{sec Simulation} and \ref{sec real}.

\section{Simulation Studies}\label{sec Simulation}
In this section, we conduct numerical experiments on a wide range of simulated data to show the effectiveness of the proposed distributed algorithms. Our implementation of the distributed algorithm is in R, to compare both the estimation accuracy and computation cost of our distributed methods with standard SIR, SAVE and DR on the full sample (denoted as the global methods).

\subsection{Simulated Data}
\subsubsection{Data Generation}
We begin with generating the predictor vector $\bm{X}$ from a multivariate normal distribution $N_p(\bm{\mu}, \bm{\Sigma})$, where $\bm{\mu}$ is the mean vector, $\bm{\Sigma}$ is the covariance matrix, and $p$ is the dimension of $\bm{X}$. Taking data magnitude and sample correlation into account, we set $\mu$ and $\bf{\Sigma}$ in the following three ways:
\begin{itemize}
    \item \textbf{Standard Normal.} Predictors are independently identically distributed, by setting $\bm{\mu} = \bm{0}_{1\times p}$ and $\bm{\Sigma} = \bm{I}_{p\times p}$.
    \item \textbf{Heterogeneous Normal.} Predictors are independently distributed, but with non-zero and unequal means, by setting $\bm{\mu}$ as $n/5$ repetitions of $\{1,2,3,4,5\}$ and $\bm{\Sigma} = \bm{I}_{p\times p}$.
    \item \textbf{Dependent Normal.} Predictors are correlated, by setting $\bm{\mu} = \bm{0}_{1\times p}$ and
$$\bm{\Sigma} = 
\left\{
\begin{aligned}
&0.8 &i = j\\
&0.5^{|i-j|} &i \neq j    
\end{aligned}.
\right.$$
The covariance matrix depicts a circumstance that correlation decreases as predictors moving away from each other. This case focuses on the influence caused by predictor dependency.
\end{itemize}

To more pertinently compare the global and distributed methods, we then generate the response $Y$ from models that the global approaches have proved to be useful. For SIR, we employ the following three models:
\begin{align*} 
\rm{Model~1} &:  y = x_1 + x_2 + x_3 + x_4  +\sigma \cdot \epsilon\\
\rm{Model~2} &:  y = (x_1+x_2)+\exp(x_3+x_4)+\sigma \cdot \epsilon\\
\rm{Model~3} &:  y = \frac{x_1}{0.5 + (x_2+1.5)^2} + \sigma \cdot \epsilon,
\end{align*}
where $\epsilon\sim N(0,1)$ and $\sigma = 0.5$.

For SAVE, we generate $Y$ based on Model 1 and the following two models:
\begin{align*} 
\rm{Model~4} &:  y = (\sqrt{2}x_1+\sqrt{2}x_2)^2+\sigma \cdot \epsilon\\
\rm{Model~5} &:  y = (x_1+x_2+1)^2+\sigma \cdot \epsilon.
\end{align*}

For DR, we generate $Y$ based on the following three models:
\begin{align*} 
\rm{Model~6} &:  y = 0.4(x_1+x_2+x_3)^2+|x_1+x_5+3x_6|^{1/2}+\sigma \cdot \epsilon\\ 
\rm{Model~7} &:  y = 0.3\sin[(x_1+x_5+3x_6)/4]+[1+(x_1+x_2+x_3)^2]\sigma \cdot \epsilon\\
\rm{Model~8} &:  y = 0.4(x_1+x_2+x_3)^2+3\sin[(x_1+x_5+3x_6)/4]+\sigma \cdot \epsilon.
\end{align*}

The true SDR directions of each model are as follows:
\begin{itemize}
    \item Model 1: $\bm{\beta}=(\frac{1}{2},\frac{1}{2},\frac{1}{2},\frac{1}{2},0,...,0)\in\mathbb{R}^p$.
    \item Model 2: $\bm{\beta}_1=(0,0,\frac{1}{\sqrt{2}},\frac{1}{\sqrt{2}},0,...,0)\in\mathbb{R}^p$, $\bm{\beta}_2=(\frac{1}{\sqrt{2}},\frac{1}{\sqrt{2}},0,...,0)\in\mathbb{R}^p$.
    \item Model 3: $\bm{\beta}_1=(1,0,...,0)\in\mathbb{R}^p$, $\bm{\beta}_2=(0,1,0,...,0)\in\mathbb{R}^p$.
    \item Model 4 and 5: $\bm{\beta}=(\frac{1}{\sqrt{2}},\frac{1}{\sqrt{2}},0,...,0)\in\mathbb{R}^p$.
    \item Model 6, 7 and 8: $\bm{\beta}_1=(\frac{1}{\sqrt{3}},\frac{1}{\sqrt{3}},\frac{1}{\sqrt{3}},0,...,0)\in\mathbb{R}^p$, $\bm{\beta}_2=(\frac{1}{\sqrt{11}},0,0,0,\frac{1}{\sqrt{11}},\frac{3}{\sqrt{11}},0,...,0)\in\mathbb{R}^p$.
\end{itemize}

In this way, a total of n observations $\{(y_i,\bm{x}_i),i=1,...,n\}$ are generated from each model.

\subsubsection{Data Partition}

In the distributed system, we scatter these observations on 5 nodes. We further consider different data partition patterns on these nodes as follows: 

\begin{itemize}
    \item \textbf{Homogeneous Equally.} We randomly partition these $n$ observations into 5 subsets with equal size. That is to say, samples are homogeneous and balanced on different workers. 
    \item \textbf{Heterogeneous Equally.} Data points are sorted according to their response values in descending order, and partitioned to each worker with equal sample size. Data magnitude gradually decreases from the first worker to the last. That is to simulate heterogeneous samples with equal size on different workers. 
    \item \textbf{Heterogeneous Unequally.} Data points are sorted according to their response values in descending order, and partitioned to each worker with unequal sample size. Data magnitude gradually decreases from the first worker to the last. The sample size on each worker is $(5\%,30\%,10\%,40\%,15\%)$ of $n$. That is to simulate an even unbalanced pattern where samples are heterogeneous on different workers with different sample size. 
\end{itemize}

\begin{figure}[H]
\centering
\includegraphics[scale=0.5]{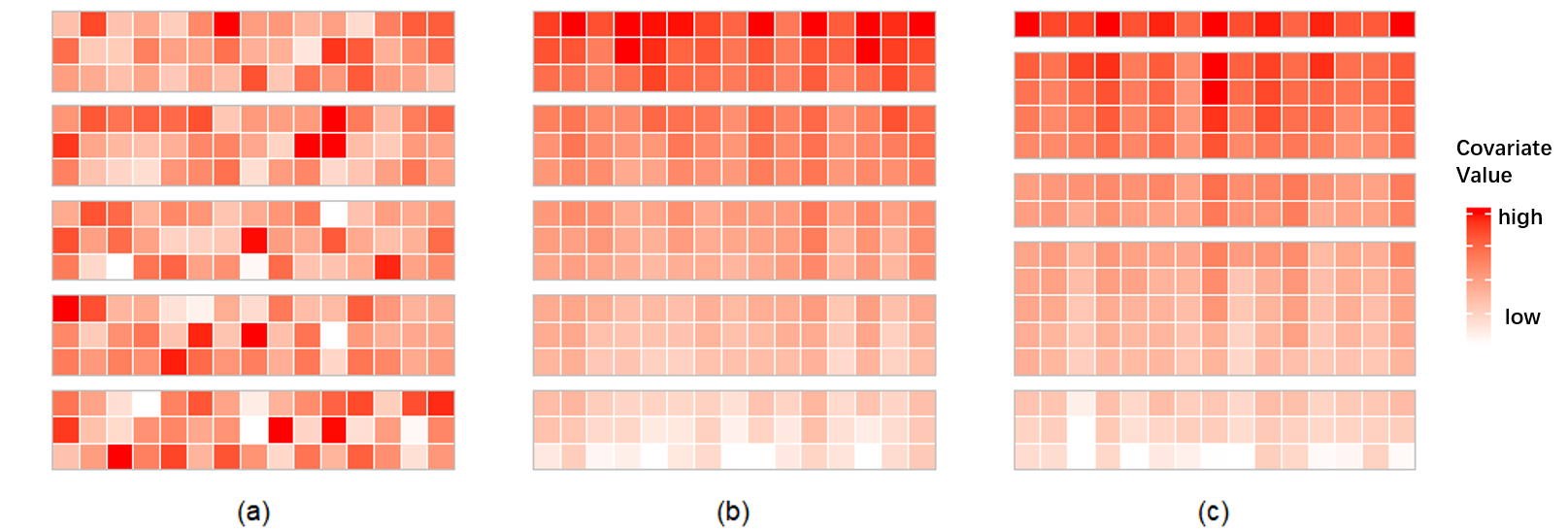}
\caption{\small Different data partition patterns. (a) Homogeneous samples of equal size on each node. (b) Heterogeneous samples of equal size on different nodes. (c) Heterogeneous samples of different sizes on different nodes.}
\label{fig partition}
\end{figure}

Figure~\ref{fig partition} shows different allocation patterns of samples on worker nodes, where each row represents an observation and each column represents a variable. The color represents the magnitude of data point. 

\subsection{Evaluation Metrics}\label{sec evalue metric}
In this article, We adopt two criteria to measure the estimation accuracy of SDR directions. The first is the trace correlation proposed by \citet{ferre1998determining}, which is defined as
\begin{equation*}
    r(K) = \frac{1}{K} \text{Tr}(\bm{P_B P_{\hat{B}}}),
\end{equation*}
where $K$ is a given dimension, $\bm{B}$ is a $p \times K$ matrix whose columns span $\mathcal{S}_{Y|X}$ and $\hat{\bm{B}}$ is its estimator. Given a full rank matrix $\bm{A}$, $\text{Tr}(\bm{A})$ is the trace of $\bm{A}$ and $\bm{P_A}$ is the projection operator in the standard inner product of $\bm{A}$, denoted by
$$\bm{P_A} = \bm{A}(\bm{A}^T \bm{A})^{-1} \bm{A}^T.$$
It can be verified that $r(K)$ ranges in $[0, 1]$, and the larger value indicates the better performance.

The second criterion is the squared multiple correlation coefficient proposed by \cite{li1991sliced}, which is defined as
\begin{equation*}
    R^2(\hat{\bm{\beta}}) = \mathop{\max}_{\bm{\beta} \in \bm{B}} \frac{(\hat{\bm{\beta}}^T \bm{\Sigma} \bm{\beta})^2}{\hat{\bm{\beta}}^T \bm{\Sigma} \hat{\bm{\beta}} \cdot \bm{\beta}^T \bm{\Sigma} \bm{\beta}},
\end{equation*}
where $\bm{\Sigma}$ is the covariance matrix of $\bm{X}$. $\hat{\bm{\beta}}$ is the estimated SDR direction, $\bm{\beta}$ is the true SDR direction and $\bm{B}$ is the SDR subspace. $R^2(\hat{\bm{\beta}})$ ranges in $[0, 1]$, and the larger value indicates the better performance.

These two metrics measure different aspects of sufficient dimension reduction. Trace correlation $r(K)$ evaluates the ability of $\mathcal{S}(\hat{\bm{B}})$, i.e., the span of the estimated SDR directions, to recover the true SDR subspace $\mathcal{S}(\bm{B})$. $R^2(\hat{\bm{\beta}})$ reveals the correlation between the reduced predictors $\hat{\bm{\beta}}^T\bm{X}$ and the real sufficient predictors $\bm{\beta}^T\bm{X}$.

\subsection{Results}
\subsubsection{Estimation Accuracy}
We conduct simulations for SIR, SAVE and DR respectively on different models, under different data generation modes and for different sample partition patterns, to verify the effectiveness of the distributed algorithm\footnote{The distributed algorithm here refers to the approximate distributed estimation of general moment-based inverse regression in Algorithm \ref{alg:dis SDR}, unless otherwise stated, because E-DSIR yields exactly the same estimator as the global SIR.} and meanwhile compare it with the global algorithm. The total sample size $n$ is set to be 1000, and the sample size on each node varies from 50 to 400. The predictor dimension $p$ is set to vary in $\{10, 20, 50\}$, the number of slices $H$ is set to be 10, and each simulation is replicated 200 times. We report the average and the standard deviation (in parentheses) of trace correlation over 200 repetitions in Table~\ref{trace SIR}, \ref{trace SAVE}, \ref{trace DR}, as well as the average and the standard deviation (in parentheses) of $R^2(\hat{\bm{\beta}})$ over 200 repetitions in Table~\ref{R2 SIR}, \ref{R2 SAVE}, \ref{R2 DR}.

\textit{\textbf{Performance of Global SIR.}} We first look at the performance of the global SIR algorithm in Table~\ref{trace SIR} and \ref{R2 SIR}. The standard SIR on the full sample is verified to be quite accurate in most cases, with the trace correlation and $R^2(\hat{\bm{\beta}})$ close to 1, which means that SIR works well in estimating the SDR directions. Except in Model 2, SIR behaves differently on $R^2(\hat{\bm{\beta}})$ under different generation mechanisms of $\bm{X}$.\footnote{When $\bm{X}$ is generated from the standard normal distribution $N_p(\bm{0}, \bm{I}_{p\times p})$, $e^{x_3+x_4}\approx 1+x_3+x_4$ according to its Taylor expansion at 0. In this case the estimated SDR directions $(\hat{\bm{\beta}}_1, \hat{\bm{\beta}}_2)$ is around $\left((\frac{1}{2},\frac{1}{2},\frac{1}{2},\frac{1}{2},0,...,0)^T, (\frac{1}{2},\frac{1}{2},-\frac{1}{2},-\frac{1}{2},0,...,0)^T\right)$, which is an orthonormal basis of the same space as $(\bm{\beta}_1, \bm{\beta}_2)$, leading to desirable trace correlation but biased $R^2(\hat{\bm{\beta}})$. In contrast, when $\bm{X}$ is generated from the non-standard normal distribution $N_p((1,2,3,4,...)^T, \bm{I}_{p\times p})$ or $\{x_1,x_2,x_3,x_4\}$ are highly correlated, $e^{x_3+x_4}$ is excessively dominant than $x_1+x_2$. In these two cases, SIR successfully identifies the first direction, but fails to estimate the second direction, leading to perfect $R^2(\hat{\bm{\beta}})$ (exactly equal to 1) but biased trace correlation.} 
In Model 3, SIR deteriorates a little for $R^2(\hat{\bm{\beta}})$ when $\bm{X}$ is generated from the heterogeneous normal and dependent normal distribution. Although the estimation in a single direction is biased, the estimated SDR space is still very close to the truth. 

\begin{figure}[H]
\centering
\includegraphics[scale=0.7]{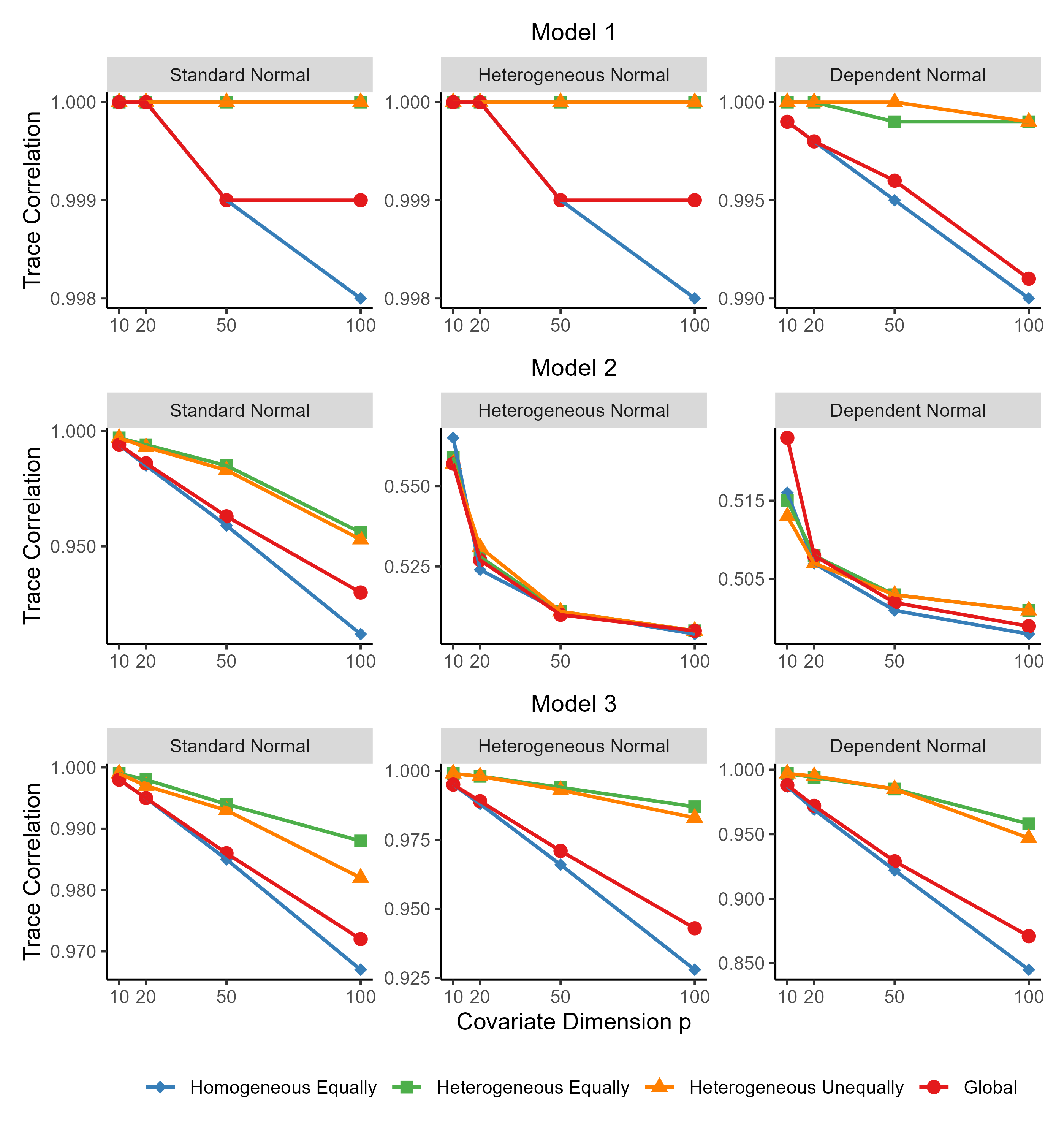}
\caption{\small Trace correlation of distributed and global SIR.}
\label{fig trace SIR}
\end{figure}

\textit{\textbf{Performance of Distributed SIR.}} Then we take a closer look at the performance of the distributed SIR algorithm. To make the comparison more intuitive, results from Table~\ref{trace SIR} and \ref{trace SIR p=100} are depicted in Figure~\ref{fig trace SIR}. It is obvious in Figure~\ref{fig trace SIR}, Table~\ref{trace SIR} and \ref{R2 SIR} that when observations are homogeneous equally scattered on worker nodes, the distributed estimator performs almost identically as the global estimator, except for the extreme condition when $p=100$ (which is discussed separately in Section \ref{sec sensitivity}). \textbf{When observations are heterogeneously scattered on different nodes, the distributed estimates are even better than the global estimate}, which is in line with our anticipation in Section \ref{sec algorithm2}. Even when $p=50$, the distributed results are still stable, although the sample size on some workers is as few as 50, only slightly larger than the predictor dimension. In general, the proposed distributed estimator is comparable to the global estimator in estimating $\mathcal{S}_{Y|X}$, even if the samples on different nodes are heterogeneous and unbalanced.

\begin{figure}[H]
\centering
\includegraphics[scale=0.7]{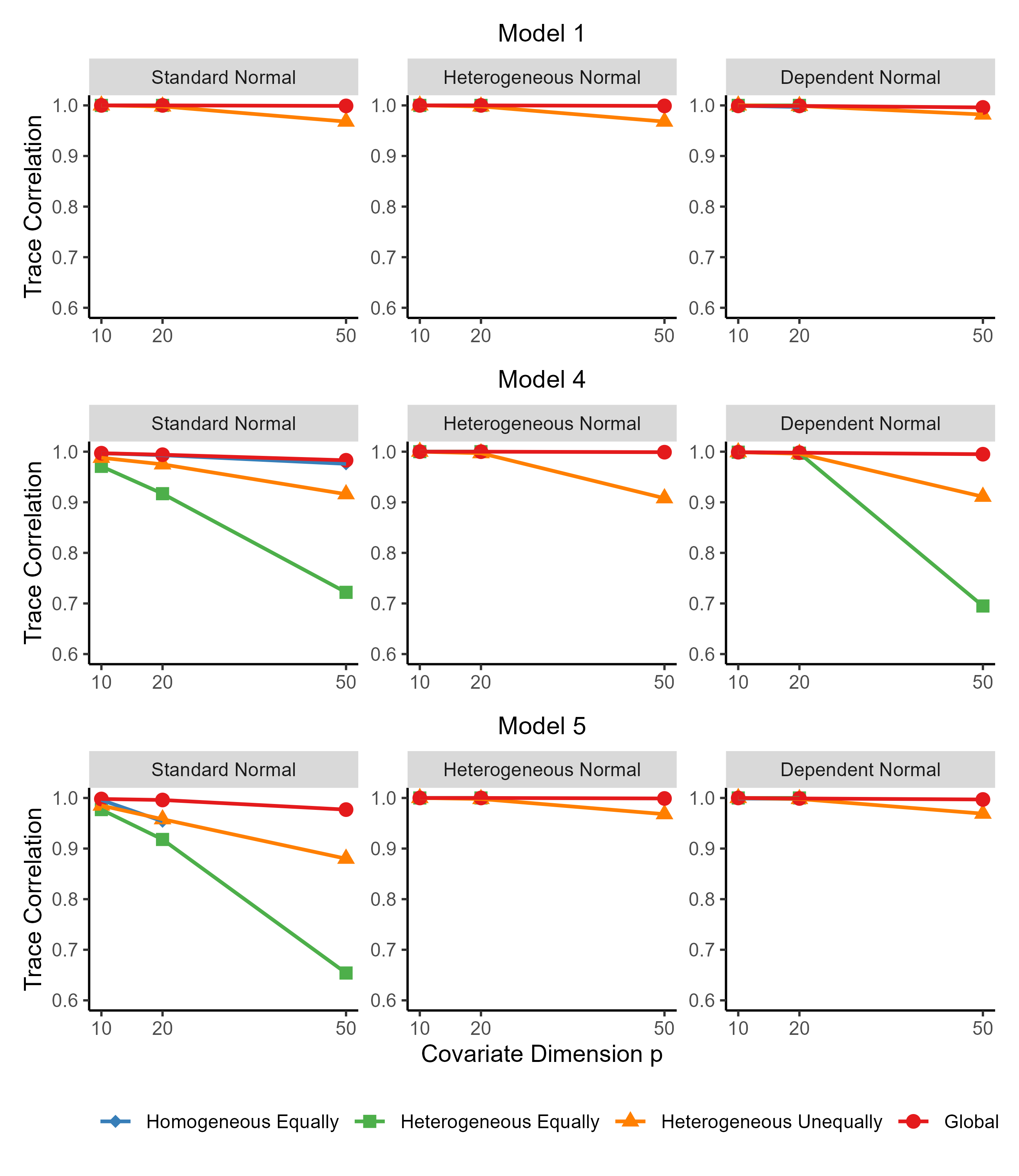}
\caption{\small Trace correlation of distributed and global SAVE.}
\label{fig trace SAVE}
\end{figure}

\begin{figure}[H]
\centering
\includegraphics[scale=0.7]{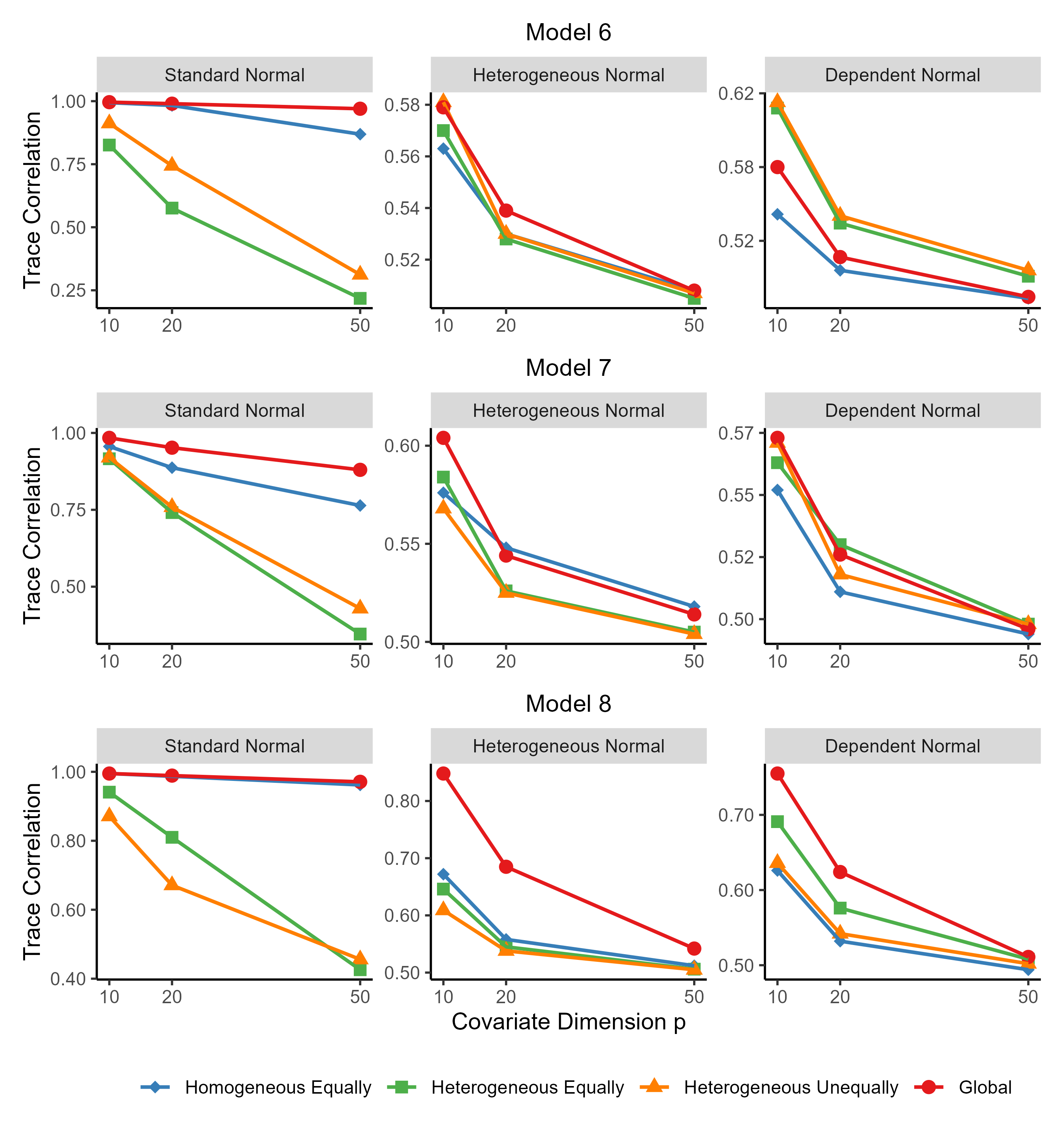}
\caption{\small Trace correlation of distributed and global DR.}
\label{fig trace DR}
\end{figure}

\textit{\textbf{Performance of Distributed SAVE and DR.}} Similar to SIR models, in SAVE models (shown in Table~\ref{trace SAVE}, \ref{R2 SAVE} and Figure~\ref{fig trace SAVE}) and DR models (shown in Table~\ref{trace DR}, \ref{R2 DR} and Figure~\ref{fig trace DR}), the distributed estimation results are generally favorable. Except on rare occasions, when $p=50$, the distributed estimation under some sample partition patterns is relatively poor, since that the sample size on individual worker node is only slightly larger than the predictor dimension. However, considering that the sample size is usually much larger than the dimension on real distributed systems, our method is still expected to be effective in most practical applications. Therefore, we conclude that the proposed distributed estimation of conditional-moment-based inverse regression preserves the statistical efficiency of the global estimation, which is applicable to homogeneous and heterogeneous, balanced and unbalanced sample partition patterns.

\begin{table}[H]
\setlength\tabcolsep{4pt}
\scriptsize
\caption{\small Average (with standard deviation) of trace correlation for distributed and global SIR algorithms.}
\label{trace SIR}
\begin{tabular}{ccccccccccccccc}
\hline
 &
  \multicolumn{4}{c}{Standard Normal} &
   &
  \multicolumn{4}{c}{Heterogeneous Normal} &
   &
  \multicolumn{4}{c}{Dependent Normal} \\ \cline{2-5} \cline{7-10} \cline{12-15} 
 &
  \begin{tabular}[c]{@{}c@{}}Homo.\\ Equal.\end{tabular} &
  \begin{tabular}[c]{@{}c@{}}Hetero.\\ Equal.\end{tabular} &
  \begin{tabular}[c]{@{}c@{}}Hetero.\\ Unequal.\end{tabular} &
  \textbf{\emph{Global}} &
   &
  \begin{tabular}[c]{@{}c@{}}Homo.\\ Equal.\end{tabular} &
  \begin{tabular}[c]{@{}c@{}}Hetero.\\ Equal.\end{tabular} &
  \begin{tabular}[c]{@{}c@{}}Hetero.\\ Unequal.\end{tabular} &
  \textbf{\emph{Global}} &
   &
  \begin{tabular}[c]{@{}c@{}}Homo.\\ Equal.\end{tabular} &
  \begin{tabular}[c]{@{}c@{}}Hetero.\\ Equal.\end{tabular} &
  \begin{tabular}[c]{@{}c@{}}Hetero.\\ Unequal.\end{tabular} &
  \textbf{\emph{Global}}  \\ \hline
Model1 &
   &
   &
   &
   &
   &
   &
   &
   &
   &
   &
   &
   &
   &
   \\
p=10 &
  \begin{tabular}[c]{@{}c@{}}1\\ (0)\end{tabular} &
  \begin{tabular}[c]{@{}c@{}}1\\ (0)\end{tabular} &
  \begin{tabular}[c]{@{}c@{}}1\\ (0)\end{tabular} &
  \begin{tabular}[c]{@{}c@{}}1\\ (0)\end{tabular} &
   &
  \begin{tabular}[c]{@{}c@{}}1\\ (0)\end{tabular} &
  \begin{tabular}[c]{@{}c@{}}1\\ (0)\end{tabular} &
  \begin{tabular}[c]{@{}c@{}}1\\ (0)\end{tabular} &
  \begin{tabular}[c]{@{}c@{}}1\\ (0)\end{tabular} &
   &
  \begin{tabular}[c]{@{}c@{}}0.999\\ (0.001)\end{tabular} &
  \begin{tabular}[c]{@{}c@{}}1\\ (0)\end{tabular} &
  \begin{tabular}[c]{@{}c@{}}1\\ (0)\end{tabular} &
  \begin{tabular}[c]{@{}c@{}}0.999\\ (0)\end{tabular} \\
p=20 &
  \begin{tabular}[c]{@{}c@{}}1\\ (0)\end{tabular} &
  \begin{tabular}[c]{@{}c@{}}1\\ (0)\end{tabular} &
  \begin{tabular}[c]{@{}c@{}}1\\ (0)\end{tabular} &
  \begin{tabular}[c]{@{}c@{}}1\\ (0)\end{tabular} &
   &
  \begin{tabular}[c]{@{}c@{}}1\\ (0)\end{tabular} &
  \begin{tabular}[c]{@{}c@{}}1\\ (0)\end{tabular} &
  \begin{tabular}[c]{@{}c@{}}1\\ (0)\end{tabular} &
  \begin{tabular}[c]{@{}c@{}}1\\ (0)\end{tabular} &
   &
  \begin{tabular}[c]{@{}c@{}}0.998\\ (0.001)\end{tabular} &
  \begin{tabular}[c]{@{}c@{}}1\\ (0)\end{tabular} &
  \begin{tabular}[c]{@{}c@{}}1\\ (0)\end{tabular} &
  \begin{tabular}[c]{@{}c@{}}0.998\\ (0.001)\end{tabular} \\
p=50 &
  \begin{tabular}[c]{@{}c@{}}0.999\\ (0)\end{tabular} &
  \begin{tabular}[c]{@{}c@{}}1\\ (0)\end{tabular} &
  \begin{tabular}[c]{@{}c@{}}1\\ (0)\end{tabular} &
  \begin{tabular}[c]{@{}c@{}}0.999\\ (0)\end{tabular} &
   &
  \begin{tabular}[c]{@{}c@{}}0.999\\ (0)\end{tabular} &
  \begin{tabular}[c]{@{}c@{}}1\\ (0)\end{tabular} &
  \begin{tabular}[c]{@{}c@{}}1\\ (0)\end{tabular} &
  \begin{tabular}[c]{@{}c@{}}0.999\\ (0)\end{tabular} &
   &
  \begin{tabular}[c]{@{}c@{}}0.995\\ (0.001)\end{tabular} &
  \begin{tabular}[c]{@{}c@{}}0.999\\ (0)\end{tabular} &
  \begin{tabular}[c]{@{}c@{}}1\\ (0)\end{tabular} &
  \begin{tabular}[c]{@{}c@{}}0.996\\ (0.001)\end{tabular} \\
  \hline
Model2 &
   &
   &
   &
   &
   &
   &
   &
   &
   &
   &
   &
   &
   &
   \\
p=10 &
  \begin{tabular}[c]{@{}c@{}}0.994\\ (0.003)\end{tabular} &
  \begin{tabular}[c]{@{}c@{}}0.997\\ (0.002)\end{tabular} &
  \begin{tabular}[c]{@{}c@{}}0.997\\ (0.002)\end{tabular} &
  \begin{tabular}[c]{@{}c@{}}0.994\\ (0.003)\end{tabular} &
   &
  \begin{tabular}[c]{@{}c@{}}0.565\\ (0.074)\end{tabular} &
  \begin{tabular}[c]{@{}c@{}}0.559\\ (0.069)\end{tabular} &
  \begin{tabular}[c]{@{}c@{}}0.557\\ (0.068)\end{tabular} &
  \begin{tabular}[c]{@{}c@{}}0.557\\ (0.062)\end{tabular} &
   &
  \begin{tabular}[c]{@{}c@{}}0.516\\ (0.026)\end{tabular} &
  \begin{tabular}[c]{@{}c@{}}0.515\\ (0.02)\end{tabular} &
  \begin{tabular}[c]{@{}c@{}}0.513\\ (0.018)\end{tabular} &
  \begin{tabular}[c]{@{}c@{}}0.523\\ (0.033)\end{tabular} \\
p=20 &
  \begin{tabular}[c]{@{}c@{}}0.985\\ (0.005)\end{tabular} &
  \begin{tabular}[c]{@{}c@{}}0.994\\ (0.003)\end{tabular} &
  \begin{tabular}[c]{@{}c@{}}0.993\\ (0.003)\end{tabular} &
  \begin{tabular}[c]{@{}c@{}}0.986\\ (0.005)\end{tabular} &
   &
  \begin{tabular}[c]{@{}c@{}}0.524\\ (0.03)\end{tabular} &
  \begin{tabular}[c]{@{}c@{}}0.528\\ (0.035)\end{tabular} &
  \begin{tabular}[c]{@{}c@{}}0.531\\ (0.039)\end{tabular} &
  \begin{tabular}[c]{@{}c@{}}0.527\\ (0.035)\end{tabular} &
   &
  \begin{tabular}[c]{@{}c@{}}0.507\\ (0.01)\end{tabular} &
  \begin{tabular}[c]{@{}c@{}}0.508\\ (0.01)\end{tabular} &
  \begin{tabular}[c]{@{}c@{}}0.507\\ (0.009)\end{tabular} &
  \begin{tabular}[c]{@{}c@{}}0.508\\ (0.011)\end{tabular} \\
p=50 &
  \begin{tabular}[c]{@{}c@{}}0.959\\ (0.008)\end{tabular} &
  \begin{tabular}[c]{@{}c@{}}0.985\\ (0.006)\end{tabular} &
  \begin{tabular}[c]{@{}c@{}}0.983\\ (0.007)\end{tabular} &
  \begin{tabular}[c]{@{}c@{}}0.963\\ (0.008)\end{tabular} &
   &
  \begin{tabular}[c]{@{}c@{}}0.511\\ (0.016)\end{tabular} &
  \begin{tabular}[c]{@{}c@{}}0.511\\ (0.015)\end{tabular} &
  \begin{tabular}[c]{@{}c@{}}0.511\\ (0.015)\end{tabular} &
  \begin{tabular}[c]{@{}c@{}}0.51\\ (0.014)\end{tabular} &
   &
  \begin{tabular}[c]{@{}c@{}}0.501\\ (0.004)\end{tabular} &
  \begin{tabular}[c]{@{}c@{}}0.503\\ (0.004)\end{tabular} &
  \begin{tabular}[c]{@{}c@{}}0.503\\ (0.005)\end{tabular} &
  \begin{tabular}[c]{@{}c@{}}0.502\\ (0.005)\end{tabular} \\
  \hline
Model3 &
   &
   &
   &
   &
   &
   &
   &
   &
   &
   &
   &
   &
   &
   \\
p=10 &
  \begin{tabular}[c]{@{}c@{}}0.998\\ (0.001)\end{tabular} &
  \begin{tabular}[c]{@{}c@{}}0.999\\ (0)\end{tabular} &
  \begin{tabular}[c]{@{}c@{}}0.999\\ (0.001)\end{tabular} &
  \begin{tabular}[c]{@{}c@{}}0.998\\ (0.001)\end{tabular} &
   &
  \begin{tabular}[c]{@{}c@{}}0.995\\ (0.003)\end{tabular} &
  \begin{tabular}[c]{@{}c@{}}0.999\\ (0.001)\end{tabular} &
  \begin{tabular}[c]{@{}c@{}}0.999\\ (0.001)\end{tabular} &
  \begin{tabular}[c]{@{}c@{}}0.995\\ (0.003)\end{tabular} &
   &
  \begin{tabular}[c]{@{}c@{}}0.987\\ (0.008)\end{tabular} &
  \begin{tabular}[c]{@{}c@{}}0.997\\ (0.002)\end{tabular} &
  \begin{tabular}[c]{@{}c@{}}0.997\\ (0.002)\end{tabular} &
  \begin{tabular}[c]{@{}c@{}}0.988\\ (0.006)\end{tabular} \\
p=20 &
  \begin{tabular}[c]{@{}c@{}}0.995\\ (0.002)\end{tabular} &
  \begin{tabular}[c]{@{}c@{}}0.998\\ (0.001)\end{tabular} &
  \begin{tabular}[c]{@{}c@{}}0.997\\ (0.001)\end{tabular} &
  \begin{tabular}[c]{@{}c@{}}0.995\\ (0.001)\end{tabular} &
   &
  \begin{tabular}[c]{@{}c@{}}0.988\\ (0.004)\end{tabular} &
  \begin{tabular}[c]{@{}c@{}}0.998\\ (0.001)\end{tabular} &
  \begin{tabular}[c]{@{}c@{}}0.998\\ (0.001)\end{tabular} &
  \begin{tabular}[c]{@{}c@{}}0.989\\ (0.003)\end{tabular} &
   &
  \begin{tabular}[c]{@{}c@{}}0.969\\ (0.013)\end{tabular} &
  \begin{tabular}[c]{@{}c@{}}0.994\\ (0.003)\end{tabular} &
  \begin{tabular}[c]{@{}c@{}}0.995\\ (0.003)\end{tabular} &
  \begin{tabular}[c]{@{}c@{}}0.972\\ (0.011)\end{tabular} \\
p=50 &
  \begin{tabular}[c]{@{}c@{}}0.985\\ (0.003)\end{tabular} &
  \begin{tabular}[c]{@{}c@{}}0.994\\ (0.001)\end{tabular} &
  \begin{tabular}[c]{@{}c@{}}0.993\\ (0.002)\end{tabular} &
  \begin{tabular}[c]{@{}c@{}}0.986\\ (0.002)\end{tabular} &
   &
  \begin{tabular}[c]{@{}c@{}}0.966\\ (0.007)\end{tabular} &
  \begin{tabular}[c]{@{}c@{}}0.994\\ (0.002)\end{tabular} &
  \begin{tabular}[c]{@{}c@{}}0.993\\ (0.003)\end{tabular} &
  \begin{tabular}[c]{@{}c@{}}0.971\\ (0.006)\end{tabular} &
   &
  \begin{tabular}[c]{@{}c@{}}0.922\\ (0.018)\end{tabular} &
  \begin{tabular}[c]{@{}c@{}}0.985\\ (0.005)\end{tabular} &
  \begin{tabular}[c]{@{}c@{}}0.985\\ (0.007)\end{tabular} &
  \begin{tabular}[c]{@{}c@{}}0.929\\ (0.016)\end{tabular} \\
  \hline
\end{tabular}
\end{table}

\begin{table}[H]
\setlength\tabcolsep{4pt}
\scriptsize
\caption{\small Average (with standard deviation) of $R^2(\hat{\bm{\beta}})$ for distributed and global SIR algorithms.}
\label{R2 SIR}
\begin{tabular}{ccccccccccccccc}
\hline
 &
  \multicolumn{4}{c}{Standard Normal} &
   &
  \multicolumn{4}{c}{Heterogeneous Normal} &
   &
  \multicolumn{4}{c}{Dependent Normal} \\ \cline{2-5} \cline{7-10} \cline{12-15} 
 &
  \begin{tabular}[c]{@{}c@{}}Homo.\\ Equal.\end{tabular} &
  \begin{tabular}[c]{@{}c@{}}Hetero.\\ Equal.\end{tabular} &
  \begin{tabular}[c]{@{}c@{}}Hetero.\\ Unequal.\end{tabular} &
  \textbf{\emph{Global}} &
   &
  \begin{tabular}[c]{@{}c@{}}Homo.\\ Equal.\end{tabular} &
  \begin{tabular}[c]{@{}c@{}}Hetero.\\ Equal.\end{tabular} &
  \begin{tabular}[c]{@{}c@{}}Hetero.\\ Unequal.\end{tabular} &
  \textbf{\emph{Global}} &
   &
  \begin{tabular}[c]{@{}c@{}}Homo.\\ Equal.\end{tabular} &
  \begin{tabular}[c]{@{}c@{}}Hetero.\\ Equal.\end{tabular} &
  \begin{tabular}[c]{@{}c@{}}Hetero.\\ Unequal.\end{tabular} &
  \textbf{\emph{Global}}  \\ \hline
Model1 &
   &
   &
   &
   &
   &
   &
   &
   &
   &
   &
   &
   &
   &
   \\
p=10 &
  \begin{tabular}[c]{@{}c@{}}1\\ (0)\end{tabular} &
  \begin{tabular}[c]{@{}c@{}}1\\ (0)\end{tabular} &
  \begin{tabular}[c]{@{}c@{}}1\\ (0)\end{tabular} &
  \begin{tabular}[c]{@{}c@{}}1\\ (0)\end{tabular} &
   &
  \begin{tabular}[c]{@{}c@{}}1\\ (0)\end{tabular} &
  \begin{tabular}[c]{@{}c@{}}1\\ (0)\end{tabular} &
  \begin{tabular}[c]{@{}c@{}}1\\ (0)\end{tabular} &
  \begin{tabular}[c]{@{}c@{}}1\\ (0)\end{tabular} &
   &
  \begin{tabular}[c]{@{}c@{}}1\\ (0)\end{tabular} &
  \begin{tabular}[c]{@{}c@{}}1\\ (0)\end{tabular} &
  \begin{tabular}[c]{@{}c@{}}1\\ (0)\end{tabular} &
  \begin{tabular}[c]{@{}c@{}}1\\ (0)\end{tabular} \\
p=20 &
  \begin{tabular}[c]{@{}c@{}}1\\ (0)\end{tabular} &
  \begin{tabular}[c]{@{}c@{}}1\\ (0)\end{tabular} &
  \begin{tabular}[c]{@{}c@{}}1\\ (0)\end{tabular} &
  \begin{tabular}[c]{@{}c@{}}1\\ (0)\end{tabular} &
   &
  \begin{tabular}[c]{@{}c@{}}1\\ (0)\end{tabular} &
  \begin{tabular}[c]{@{}c@{}}1\\ (0)\end{tabular} &
  \begin{tabular}[c]{@{}c@{}}1\\ (0)\end{tabular} &
  \begin{tabular}[c]{@{}c@{}}1\\ (0)\end{tabular} &
   &
  \begin{tabular}[c]{@{}c@{}}1\\ (0)\end{tabular} &
  \begin{tabular}[c]{@{}c@{}}1\\ (0)\end{tabular} &
  \begin{tabular}[c]{@{}c@{}}1\\ (0)\end{tabular} &
  \begin{tabular}[c]{@{}c@{}}1\\ (0)\end{tabular} \\
p=50 &
  \begin{tabular}[c]{@{}c@{}}0.999\\ (0)\end{tabular} &
  \begin{tabular}[c]{@{}c@{}}1\\ (0)\end{tabular} &
  \begin{tabular}[c]{@{}c@{}}1\\ (0)\end{tabular} &
  \begin{tabular}[c]{@{}c@{}}0.999\\ (0)\end{tabular} &
   &
  \begin{tabular}[c]{@{}c@{}}0.999\\ (0)\end{tabular} &
  \begin{tabular}[c]{@{}c@{}}1\\ (0)\end{tabular} &
  \begin{tabular}[c]{@{}c@{}}1\\ (0)\end{tabular} &
  \begin{tabular}[c]{@{}c@{}}0.999\\ (0)\end{tabular} &
   &
  \begin{tabular}[c]{@{}c@{}}0.999\\ (0)\end{tabular} &
  \begin{tabular}[c]{@{}c@{}}1\\ (0)\end{tabular} &
  \begin{tabular}[c]{@{}c@{}}1\\ (0)\end{tabular} &
  \begin{tabular}[c]{@{}c@{}}0.999\\ (0)\end{tabular} \\ \hline
Model2 &
   &
   &
   &
   &
   &
   &
   &
   &
   &
   &
   &
   &
   &
   \\
p=10 &
  \begin{tabular}[c]{@{}c@{}}0.58\\ (0.017)\end{tabular} &
  \begin{tabular}[c]{@{}c@{}}0.523\\ (0.01)\end{tabular} &
  \begin{tabular}[c]{@{}c@{}}0.53\\ (0.012)\end{tabular} &
  \begin{tabular}[c]{@{}c@{}}0.584\\ (0.014)\end{tabular} &
   &
  \begin{tabular}[c]{@{}c@{}}1\\ (0)\end{tabular} &
  \begin{tabular}[c]{@{}c@{}}1\\ (0)\end{tabular} &
  \begin{tabular}[c]{@{}c@{}}1\\ (0)\end{tabular} &
  \begin{tabular}[c]{@{}c@{}}1\\ (0)\end{tabular} &
   &
  \begin{tabular}[c]{@{}c@{}}1\\ (0)\end{tabular} &
  \begin{tabular}[c]{@{}c@{}}1\\ (0)\end{tabular} &
  \begin{tabular}[c]{@{}c@{}}1\\ (0)\end{tabular} &
  \begin{tabular}[c]{@{}c@{}}1\\ (0)\end{tabular} \\
p=20 &
  \begin{tabular}[c]{@{}c@{}}0.592\\ (0.016)\end{tabular} &
  \begin{tabular}[c]{@{}c@{}}0.516\\ (0.01)\end{tabular} &
  \begin{tabular}[c]{@{}c@{}}0.518\\ (0.011)\end{tabular} &
  \begin{tabular}[c]{@{}c@{}}0.573\\ (0.014)\end{tabular} &
   &
  \begin{tabular}[c]{@{}c@{}}1\\ (0)\end{tabular} &
  \begin{tabular}[c]{@{}c@{}}1\\ (0)\end{tabular} &
  \begin{tabular}[c]{@{}c@{}}1\\ (0)\end{tabular} &
  \begin{tabular}[c]{@{}c@{}}1\\ (0)\end{tabular} &
   &
  \begin{tabular}[c]{@{}c@{}}1\\ (0)\end{tabular} &
  \begin{tabular}[c]{@{}c@{}}1\\ (0)\end{tabular} &
  \begin{tabular}[c]{@{}c@{}}1\\ (0)\end{tabular} &
  \begin{tabular}[c]{@{}c@{}}1\\ (0)\end{tabular} \\
p=50 &
  \begin{tabular}[c]{@{}c@{}}0.569\\ (0.017)\end{tabular} &
  \begin{tabular}[c]{@{}c@{}}0.537\\ (0.015)\end{tabular} &
  \begin{tabular}[c]{@{}c@{}}0.507\\ (0.01)\end{tabular} &
  \begin{tabular}[c]{@{}c@{}}0.565\\ (0.016)\end{tabular} &
   &
  \begin{tabular}[c]{@{}c@{}}0.999\\ (0)\end{tabular} &
  \begin{tabular}[c]{@{}c@{}}1\\ (0)\end{tabular} &
  \begin{tabular}[c]{@{}c@{}}1\\ (0)\end{tabular} &
  \begin{tabular}[c]{@{}c@{}}0.999\\ (0)\end{tabular} &
   &
  \begin{tabular}[c]{@{}c@{}}0.999\\ (0)\end{tabular} &
  \begin{tabular}[c]{@{}c@{}}1\\ (0)\end{tabular} &
  \begin{tabular}[c]{@{}c@{}}1\\ (0)\end{tabular} &
  \begin{tabular}[c]{@{}c@{}}0.999\\ (0)\end{tabular} \\ \hline
Model3 &
   &
   &
   &
   &
   &
   &
   &
   &
   &
   &
   &
   &
   &
   \\
p=10 &
  \begin{tabular}[c]{@{}c@{}}0.998\\ (0.001)\end{tabular} &
  \begin{tabular}[c]{@{}c@{}}0.998\\ (0.002)\end{tabular} &
  \begin{tabular}[c]{@{}c@{}}0.998\\ (0.002)\end{tabular} &
  \begin{tabular}[c]{@{}c@{}}0.999\\ (0.001)\end{tabular} &
   &
  \begin{tabular}[c]{@{}c@{}}0.811\\ (0.012)\end{tabular} &
  \begin{tabular}[c]{@{}c@{}}0.792\\ (0.008)\end{tabular} &
  \begin{tabular}[c]{@{}c@{}}0.787\\ (0.008)\end{tabular} &
  \begin{tabular}[c]{@{}c@{}}0.803\\ (0.018)\end{tabular} &
   &
  \begin{tabular}[c]{@{}c@{}}0.862\\ (0.01)\end{tabular} &
  \begin{tabular}[c]{@{}c@{}}0.928\\ (0.004)\end{tabular} &
  \begin{tabular}[c]{@{}c@{}}0.913\\ (0.005)\end{tabular} &
  \begin{tabular}[c]{@{}c@{}}0.969\\ (0.008)\end{tabular} \\
p=20 &
  \begin{tabular}[c]{@{}c@{}}0.997\\ (0.002)\end{tabular} &
  \begin{tabular}[c]{@{}c@{}}0.998\\ (0.002)\end{tabular} &
  \begin{tabular}[c]{@{}c@{}}0.997\\ (0.002)\end{tabular} &
  \begin{tabular}[c]{@{}c@{}}0.998\\ (0.001)\end{tabular} &
   &
  \begin{tabular}[c]{@{}c@{}}0.817\\ (0.012)\end{tabular} &
  \begin{tabular}[c]{@{}c@{}}0.793\\ (0.008)\end{tabular} &
  \begin{tabular}[c]{@{}c@{}}0.783\\ (0.008)\end{tabular} &
  \begin{tabular}[c]{@{}c@{}}0.803\\ (0.009)\end{tabular} &
   &
  \begin{tabular}[c]{@{}c@{}}0.859\\ (0.011)\end{tabular} &
  \begin{tabular}[c]{@{}c@{}}0.921\\ (0.004)\end{tabular} &
  \begin{tabular}[c]{@{}c@{}}0.918\\ (0.004)\end{tabular} &
  \begin{tabular}[c]{@{}c@{}}0.96\\ (0.008)\end{tabular} \\
p=50 &
  \begin{tabular}[c]{@{}c@{}}0.994\\ (0.002)\end{tabular} &
  \begin{tabular}[c]{@{}c@{}}0.996\\ (0.002)\end{tabular} &
  \begin{tabular}[c]{@{}c@{}}0.995\\ (0.002)\end{tabular} &
  \begin{tabular}[c]{@{}c@{}}0.995\\ (0.001)\end{tabular} &
   &
  \begin{tabular}[c]{@{}c@{}}0.8\\ (0.011)\end{tabular} &
  \begin{tabular}[c]{@{}c@{}}0.791\\ (0.007)\end{tabular} &
  \begin{tabular}[c]{@{}c@{}}0.781\\ (0.008)\end{tabular} &
  \begin{tabular}[c]{@{}c@{}}0.802\\ (0.01)\end{tabular} &
   &
  \begin{tabular}[c]{@{}c@{}}0.838\\ (0.013)\end{tabular} &
  \begin{tabular}[c]{@{}c@{}}0.916\\ (0.005)\end{tabular} &
  \begin{tabular}[c]{@{}c@{}}0.913\\ (0.006)\end{tabular} &
  \begin{tabular}[c]{@{}c@{}}0.93\\ (0.012)\end{tabular} \\ \hline
\end{tabular}
\end{table}

\begin{table}[H]
\setlength\tabcolsep{4pt}
\scriptsize
\caption{\small Average (with standard deviation) of trace correlation for distributed and global SAVE algorithms.}
\label{trace SAVE}
\begin{tabular}{ccccccccccccccc}
\hline
 &
  \multicolumn{4}{c}{Standard Normal} &
   &
  \multicolumn{4}{c}{Heterogeneous Normal} &
   &
  \multicolumn{4}{c}{Dependent Normal} \\ \cline{2-5} \cline{7-10} \cline{12-15} 
 &
  \begin{tabular}[c]{@{}c@{}}Homo.\\ Equal.\end{tabular} &
  \begin{tabular}[c]{@{}c@{}}Hetero.\\ Equal.\end{tabular} &
  \begin{tabular}[c]{@{}c@{}}Hetero.\\ Unequal.\end{tabular} &
  \textbf{\emph{Global}} &
   &
  \begin{tabular}[c]{@{}c@{}}Homo.\\ Equal.\end{tabular} &
  \begin{tabular}[c]{@{}c@{}}Hetero.\\ Equal.\end{tabular} &
  \begin{tabular}[c]{@{}c@{}}Hetero.\\ Unequal.\end{tabular} &
  \textbf{\emph{Global}} &
   &
  \begin{tabular}[c]{@{}c@{}}Homo.\\ Equal.\end{tabular} &
  \begin{tabular}[c]{@{}c@{}}Hetero.\\ Equal.\end{tabular} &
  \begin{tabular}[c]{@{}c@{}}Hetero.\\ Unequal.\end{tabular} &
  \textbf{\emph{Global}}  \\ \hline
Model1 &
   &
   &
   &
   &
   &
   &
   &
   &
   &
   &
   &
   &
   &
   \\
p=10 &
  \begin{tabular}[c]{@{}c@{}}1\\ (0)\end{tabular} &
  \begin{tabular}[c]{@{}c@{}}1\\ (0)\end{tabular} &
  \begin{tabular}[c]{@{}c@{}}1\\ (0)\end{tabular} &
  \begin{tabular}[c]{@{}c@{}}1\\ (0)\end{tabular} &
   &
  \begin{tabular}[c]{@{}c@{}}1\\ (0)\end{tabular} &
  \begin{tabular}[c]{@{}c@{}}1\\ (0)\end{tabular} &
  \begin{tabular}[c]{@{}c@{}}1\\ (0)\end{tabular} &
  \begin{tabular}[c]{@{}c@{}}1\\ (0)\end{tabular} &
   &
  \begin{tabular}[c]{@{}c@{}}0.999\\ (0)\end{tabular} &
  \begin{tabular}[c]{@{}c@{}}1\\ (0)\end{tabular} &
  \begin{tabular}[c]{@{}c@{}}1\\ (0)\end{tabular} &
  \begin{tabular}[c]{@{}c@{}}0.999\\ (0)\end{tabular} \\
p=20 &
  \begin{tabular}[c]{@{}c@{}}1\\ (0)\end{tabular} &
  \begin{tabular}[c]{@{}c@{}}1\\ (0)\end{tabular} &
  \begin{tabular}[c]{@{}c@{}}0.998\\ (0.002)\end{tabular} &
  \begin{tabular}[c]{@{}c@{}}1\\ (0)\end{tabular} &
   &
  \begin{tabular}[c]{@{}c@{}}1\\ (0)\end{tabular} &
  \begin{tabular}[c]{@{}c@{}}1\\ (0)\end{tabular} &
  \begin{tabular}[c]{@{}c@{}}0.998\\ (0.002)\end{tabular} &
  \begin{tabular}[c]{@{}c@{}}1\\ (0)\end{tabular} &
   &
  \begin{tabular}[c]{@{}c@{}}0.997\\ (0.001)\end{tabular} &
  \begin{tabular}[c]{@{}c@{}}1\\ (0)\end{tabular} &
  \begin{tabular}[c]{@{}c@{}}0.999\\ (0.002)\end{tabular} &
  \begin{tabular}[c]{@{}c@{}}0.999\\ (0.001)\end{tabular} \\
p=50 &
  \begin{tabular}[c]{@{}c@{}}0.002\\ (0.003)\end{tabular} &
  \begin{tabular}[c]{@{}c@{}}0.059\\ (0.083)\end{tabular} &
  \begin{tabular}[c]{@{}c@{}}0.968\\ (0.013)\end{tabular} &
  \begin{tabular}[c]{@{}c@{}}0.999\\ (0)\end{tabular} &
   &
  \begin{tabular}[c]{@{}c@{}}0.002\\ (0.002)\end{tabular} &
  \begin{tabular}[c]{@{}c@{}}0.062\\ (0.084)\end{tabular} &
  \begin{tabular}[c]{@{}c@{}}0.968\\ (0.013)\end{tabular} &
  \begin{tabular}[c]{@{}c@{}}0.999\\ (0)\end{tabular} &
   &
  \begin{tabular}[c]{@{}c@{}}0.002\\ (0.002)\end{tabular} &
  \begin{tabular}[c]{@{}c@{}}0.007\\ (0.009)\end{tabular} &
  \begin{tabular}[c]{@{}c@{}}0.982\\ (0.013)\end{tabular} &
  \begin{tabular}[c]{@{}c@{}}0.996\\ (0.001)\end{tabular} \\ \hline
Model4 &
   &
   &
   &
   &
   &
   &
   &
   &
   &
   &
   &
   &
   &
   \\
p=10 &
  \begin{tabular}[c]{@{}c@{}}0.997\\ (0.001)\end{tabular} &
  \begin{tabular}[c]{@{}c@{}}0.971\\ (0.015)\end{tabular} &
  \begin{tabular}[c]{@{}c@{}}0.988\\ (0.005)\end{tabular} &
  \begin{tabular}[c]{@{}c@{}}0.997\\ (0.001)\end{tabular} &
   &
  \begin{tabular}[c]{@{}c@{}}1\\ (0)\end{tabular} &
  \begin{tabular}[c]{@{}c@{}}1\\ (0)\end{tabular} &
  \begin{tabular}[c]{@{}c@{}}1\\ (0)\end{tabular} &
  \begin{tabular}[c]{@{}c@{}}1\\ (0)\end{tabular} &
   &
  \begin{tabular}[c]{@{}c@{}}0.999\\ (0.001)\end{tabular} &
  \begin{tabular}[c]{@{}c@{}}0.999\\ (0)\end{tabular} &
  \begin{tabular}[c]{@{}c@{}}0.999\\ (0)\end{tabular} &
  \begin{tabular}[c]{@{}c@{}}0.999\\ (0)\end{tabular} \\
p=20 &
  \begin{tabular}[c]{@{}c@{}}0.993\\ (0.002)\end{tabular} &
  \begin{tabular}[c]{@{}c@{}}0.917\\ (0.039)\end{tabular} &
  \begin{tabular}[c]{@{}c@{}}0.975\\ (0.011)\end{tabular} &
  \begin{tabular}[c]{@{}c@{}}0.994\\ (0.002)\end{tabular} &
   &
  \begin{tabular}[c]{@{}c@{}}1\\ (0)\end{tabular} &
  \begin{tabular}[c]{@{}c@{}}1\\ (0)\end{tabular} &
  \begin{tabular}[c]{@{}c@{}}0.997\\ (0.002)\end{tabular} &
  \begin{tabular}[c]{@{}c@{}}1\\ (0)\end{tabular} &
   &
  \begin{tabular}[c]{@{}c@{}}0.997\\ (0.001)\end{tabular} &
  \begin{tabular}[c]{@{}c@{}}0.997\\ (0.002)\end{tabular} &
  \begin{tabular}[c]{@{}c@{}}0.996\\ (0.003)\end{tabular} &
  \begin{tabular}[c]{@{}c@{}}0.998\\ (0.001)\end{tabular} \\
p=50 &
  \begin{tabular}[c]{@{}c@{}}0.976\\ (0.005)\end{tabular} &
  \begin{tabular}[c]{@{}c@{}}0.722\\ (0.145)\end{tabular} &
  \begin{tabular}[c]{@{}c@{}}0.916\\ (0.031)\end{tabular} &
  \begin{tabular}[c]{@{}c@{}}0.983\\ (0.003)\end{tabular} &
   &
  \begin{tabular}[c]{@{}c@{}}0.002\\ (0.003)\end{tabular} &
  \begin{tabular}[c]{@{}c@{}}0.498\\ (0.38)\end{tabular} &
  \begin{tabular}[c]{@{}c@{}}0.908\\ (0.058)\end{tabular} &
  \begin{tabular}[c]{@{}c@{}}0.999\\ (0)\end{tabular} &
   &
  \begin{tabular}[c]{@{}c@{}}0.003\\ (0.004)\end{tabular} &
  \begin{tabular}[c]{@{}c@{}}0.695\\ (0.368)\end{tabular} &
  \begin{tabular}[c]{@{}c@{}}0.911\\ (0.088)\end{tabular} &
  \begin{tabular}[c]{@{}c@{}}0.995\\ (0.002)\end{tabular} \\ \hline
Model5 &
   &
   &
   &
   &
   &
   &
   &
   &
   &
   &
   &
   &
   &
   \\
p=10 &
  \begin{tabular}[c]{@{}c@{}}0.996\\ (0.002)\end{tabular} &
  \begin{tabular}[c]{@{}c@{}}0.977\\ (0.014)\end{tabular} &
  \begin{tabular}[c]{@{}c@{}}0.985\\ (0.01)\end{tabular} &
  \begin{tabular}[c]{@{}c@{}}0.998\\ (0.001)\end{tabular} &
   &
  \begin{tabular}[c]{@{}c@{}}1\\ (0)\end{tabular} &
  \begin{tabular}[c]{@{}c@{}}1\\ (0)\end{tabular} &
  \begin{tabular}[c]{@{}c@{}}1\\ (0)\end{tabular} &
  \begin{tabular}[c]{@{}c@{}}1\\ (0)\end{tabular} &
   &
  \begin{tabular}[c]{@{}c@{}}0.999\\ (0)\end{tabular} &
  \begin{tabular}[c]{@{}c@{}}1\\ (0)\end{tabular} &
  \begin{tabular}[c]{@{}c@{}}1\\ (0)\end{tabular} &
  \begin{tabular}[c]{@{}c@{}}1\\ (0)\end{tabular} \\
p=20 &
  \begin{tabular}[c]{@{}c@{}}0.954\\ (0.022)\end{tabular} &
  \begin{tabular}[c]{@{}c@{}}0.918\\ (0.039)\end{tabular} &
  \begin{tabular}[c]{@{}c@{}}0.958\\ (0.023)\end{tabular} &
  \begin{tabular}[c]{@{}c@{}}0.996\\ (0.002)\end{tabular} &
   &
  \begin{tabular}[c]{@{}c@{}}1\\ (0)\end{tabular} &
  \begin{tabular}[c]{@{}c@{}}1\\ (0)\end{tabular} &
  \begin{tabular}[c]{@{}c@{}}0.998\\ (0.002)\end{tabular} &
  \begin{tabular}[c]{@{}c@{}}1\\ (0)\end{tabular} &
   &
  \begin{tabular}[c]{@{}c@{}}0.998\\ (0.001)\end{tabular} &
  \begin{tabular}[c]{@{}c@{}}1\\ (0)\end{tabular} &
  \begin{tabular}[c]{@{}c@{}}0.998\\ (0.002)\end{tabular} &
  \begin{tabular}[c]{@{}c@{}}0.999\\ (0)\end{tabular} \\
p=50 &
  \begin{tabular}[c]{@{}c@{}}0.02\\ (0.043)\end{tabular} &
  \begin{tabular}[c]{@{}c@{}}0.654\\ (0.159)\end{tabular} &
  \begin{tabular}[c]{@{}c@{}}0.88\\ (0.084)\end{tabular} &
  \begin{tabular}[c]{@{}c@{}}0.977\\ (0.006)\end{tabular} &
   &
  \begin{tabular}[c]{@{}c@{}}0.002\\ (0.003)\end{tabular} &
  \begin{tabular}[c]{@{}c@{}}0.052\\ (0.087)\end{tabular} &
  \begin{tabular}[c]{@{}c@{}}0.968\\ (0.013)\end{tabular} &
  \begin{tabular}[c]{@{}c@{}}0.999\\ (0)\end{tabular} &
   &
  \begin{tabular}[c]{@{}c@{}}0.002\\ (0.003)\end{tabular} &
  \begin{tabular}[c]{@{}c@{}}0.044\\ (0.166)\end{tabular} &
  \begin{tabular}[c]{@{}c@{}}0.969\\ (0.02)\end{tabular} &
  \begin{tabular}[c]{@{}c@{}}0.997\\ (0.001)\end{tabular} \\ \hline
\end{tabular}
\end{table}

\begin{table}[H]
\setlength\tabcolsep{4pt}
\scriptsize
\caption{\small Average (with standard deviation) of trace correlation for distributed and global DR algorithms.}
\label{trace DR}
\begin{tabular}{ccccccccccccccc}
\hline
 &
  \multicolumn{4}{c}{Standard Normal} &
   &
  \multicolumn{4}{c}{Heterogeneous Normal} &
   &
  \multicolumn{4}{c}{Dependent Normal} \\ \cline{2-5} \cline{7-10} \cline{12-15} 
 &
  \begin{tabular}[c]{@{}c@{}}Homo.\\ Equal.\end{tabular} &
  \begin{tabular}[c]{@{}c@{}}Hetero.\\ Equal.\end{tabular} &
  \begin{tabular}[c]{@{}c@{}}Hetero.\\ Unequal.\end{tabular} &
  \textbf{\emph{Global}} &
   &
  \begin{tabular}[c]{@{}c@{}}Homo.\\ Equal.\end{tabular} &
  \begin{tabular}[c]{@{}c@{}}Hetero.\\ Equal.\end{tabular} &
  \begin{tabular}[c]{@{}c@{}}Hetero.\\ Unequal.\end{tabular} &
  \textbf{\emph{Global}} &
   &
  \begin{tabular}[c]{@{}c@{}}Homo.\\ Equal.\end{tabular} &
  \begin{tabular}[c]{@{}c@{}}Hetero.\\ Equal.\end{tabular} &
  \begin{tabular}[c]{@{}c@{}}Hetero.\\ Unequal.\end{tabular} &
  \textbf{\emph{Global}}  \\ \hline
Model6 &
   &
   &
   &
   &
   &
   &
   &
   &
   &
   &
   &
   &
   &
   \\
p=10 &
  \begin{tabular}[c]{@{}c@{}}0.994\\ (0.002)\end{tabular} &
  \begin{tabular}[c]{@{}c@{}}0.826\\ (0.098)\end{tabular} &
  \begin{tabular}[c]{@{}c@{}}0.912\\ (0.051)\end{tabular} &
  \begin{tabular}[c]{@{}c@{}}0.996\\ (0.001)\end{tabular} &
   &
  \begin{tabular}[c]{@{}c@{}}0.563\\ (0.072)\end{tabular} &
  \begin{tabular}[c]{@{}c@{}}0.57\\ (0.08)\end{tabular} &
  \begin{tabular}[c]{@{}c@{}}0.581\\ (0.093)\end{tabular} &
  \begin{tabular}[c]{@{}c@{}}0.579\\ (0.084)\end{tabular} &
   &
  \begin{tabular}[c]{@{}c@{}}0.543\\ (0.078)\end{tabular} &
  \begin{tabular}[c]{@{}c@{}}0.615\\ (0.105)\end{tabular} &
  \begin{tabular}[c]{@{}c@{}}0.619\\ (0.107)\end{tabular} &
  \begin{tabular}[c]{@{}c@{}}0.575\\ (0.096)\end{tabular} \\
p=20 &
  \begin{tabular}[c]{@{}c@{}}0.983\\ (0.004)\end{tabular} &
  \begin{tabular}[c]{@{}c@{}}0.576\\ (0.111)\end{tabular} &
  \begin{tabular}[c]{@{}c@{}}0.744\\ (0.099)\end{tabular} &
  \begin{tabular}[c]{@{}c@{}}0.99\\ (0.002)\end{tabular} &
   &
  \begin{tabular}[c]{@{}c@{}}0.53\\ (0.038)\end{tabular} &
  \begin{tabular}[c]{@{}c@{}}0.528\\ (0.038)\end{tabular} &
  \begin{tabular}[c]{@{}c@{}}0.53\\ (0.04)\end{tabular} &
  \begin{tabular}[c]{@{}c@{}}0.539\\ (0.047)\end{tabular} &
   &
  \begin{tabular}[c]{@{}c@{}}0.505\\ (0.035)\end{tabular} &
  \begin{tabular}[c]{@{}c@{}}0.537\\ (0.055)\end{tabular} &
  \begin{tabular}[c]{@{}c@{}}0.542\\ (0.057)\end{tabular} &
  \begin{tabular}[c]{@{}c@{}}0.514\\ (0.049)\end{tabular} \\
p=50 &
  \begin{tabular}[c]{@{}c@{}}0.869\\ (0.045)\end{tabular} &
  \begin{tabular}[c]{@{}c@{}}0.218\\ (0.084)\end{tabular} &
  \begin{tabular}[c]{@{}c@{}}0.312\\ (0.1)\end{tabular} &
  \begin{tabular}[c]{@{}c@{}}0.97\\ (0.005)\end{tabular} &
   &
  \begin{tabular}[c]{@{}c@{}}0.508\\ (0.015)\end{tabular} &
  \begin{tabular}[c]{@{}c@{}}0.505\\ (0.014)\end{tabular} &
  \begin{tabular}[c]{@{}c@{}}0.507\\ (0.019)\end{tabular} &
  \begin{tabular}[c]{@{}c@{}}0.508\\ (0.015)\end{tabular} &
   &
  \begin{tabular}[c]{@{}c@{}}0.486\\ (0.016)\end{tabular} &
  \begin{tabular}[c]{@{}c@{}}0.501\\ (0.018)\end{tabular} &
  \begin{tabular}[c]{@{}c@{}}0.505\\ (0.021)\end{tabular} &
  \begin{tabular}[c]{@{}c@{}}0.487\\ (0.017)\end{tabular} \\ \hline
Model7 &
   &
   &
   &
   &
   &
   &
   &
   &
   &
   &
   &
   &
   &
   \\
p=10 &
  \begin{tabular}[c]{@{}c@{}}0.956\\ (0.085)\end{tabular} &
  \begin{tabular}[c]{@{}c@{}}0.916\\ (0.034)\end{tabular} &
  \begin{tabular}[c]{@{}c@{}}0.92\\ (0.044)\end{tabular} &
  \begin{tabular}[c]{@{}c@{}}0.984\\ (0.037)\end{tabular} &
   &
  \begin{tabular}[c]{@{}c@{}}0.576\\ (0.1)\end{tabular} &
  \begin{tabular}[c]{@{}c@{}}0.584\\ (0.105)\end{tabular} &
  \begin{tabular}[c]{@{}c@{}}0.568\\ (0.09)\end{tabular} &
  \begin{tabular}[c]{@{}c@{}}0.604\\ (0.142)\end{tabular} &
   &
  \begin{tabular}[c]{@{}c@{}}0.552\\ (0.086)\end{tabular} &
  \begin{tabular}[c]{@{}c@{}}0.563\\ (0.092)\end{tabular} &
  \begin{tabular}[c]{@{}c@{}}0.571\\ (0.098)\end{tabular} &
  \begin{tabular}[c]{@{}c@{}}0.573\\ (0.117)\end{tabular} \\
p=20 &
  \begin{tabular}[c]{@{}c@{}}0.887\\ (0.138)\end{tabular} &
  \begin{tabular}[c]{@{}c@{}}0.741\\ (0.11)\end{tabular} &
  \begin{tabular}[c]{@{}c@{}}0.759\\ (0.106)\end{tabular} &
  \begin{tabular}[c]{@{}c@{}}0.952\\ (0.067)\end{tabular} &
   &
  \begin{tabular}[c]{@{}c@{}}0.548\\ (0.087)\end{tabular} &
  \begin{tabular}[c]{@{}c@{}}0.526\\ (0.036)\end{tabular} &
  \begin{tabular}[c]{@{}c@{}}0.525\\ (0.049)\end{tabular} &
  \begin{tabular}[c]{@{}c@{}}0.544\\ (0.09)\end{tabular} &
   &
  \begin{tabular}[c]{@{}c@{}}0.511\\ (0.056)\end{tabular} &
  \begin{tabular}[c]{@{}c@{}}0.53\\ (0.049)\end{tabular} &
  \begin{tabular}[c]{@{}c@{}}0.518\\ (0.041)\end{tabular} &
  \begin{tabular}[c]{@{}c@{}}0.526\\ (0.082)\end{tabular} \\
p=50 &
  \begin{tabular}[c]{@{}c@{}}0.764\\ (0.17)\end{tabular} &
  \begin{tabular}[c]{@{}c@{}}0.346\\ (0.113)\end{tabular} &
  \begin{tabular}[c]{@{}c@{}}0.429\\ (0.101)\end{tabular} &
  \begin{tabular}[c]{@{}c@{}}0.88\\ (0.114)\end{tabular} &
   &
  \begin{tabular}[c]{@{}c@{}}0.518\\ (0.064)\end{tabular} &
  \begin{tabular}[c]{@{}c@{}}0.505\\ (0.013)\end{tabular} &
  \begin{tabular}[c]{@{}c@{}}0.504\\ (0.016)\end{tabular} &
  \begin{tabular}[c]{@{}c@{}}0.514\\ (0.049)\end{tabular} &
   &
  \begin{tabular}[c]{@{}c@{}}0.494\\ (0.05)\end{tabular} &
  \begin{tabular}[c]{@{}c@{}}0.498\\ (0.024)\end{tabular} &
  \begin{tabular}[c]{@{}c@{}}0.498\\ (0.017)\end{tabular} &
  \begin{tabular}[c]{@{}c@{}}0.496\\ (0.056)\end{tabular} \\ \hline
Model8 &
   &
   &
   &
   &
   &
   &
   &
   &
   &
   &
   &
   &
   &
   \\
p=10 &
  \begin{tabular}[c]{@{}c@{}}0.995\\ (0.002)\end{tabular} &
  \begin{tabular}[c]{@{}c@{}}0.941\\ (0.026)\end{tabular} &
  \begin{tabular}[c]{@{}c@{}}0.871\\ (0.083)\end{tabular} &
  \begin{tabular}[c]{@{}c@{}}0.995\\ (0.002)\end{tabular} &
   &
  \begin{tabular}[c]{@{}c@{}}0.672\\ (0.121)\end{tabular} &
  \begin{tabular}[c]{@{}c@{}}0.646\\ (0.119)\end{tabular} &
  \begin{tabular}[c]{@{}c@{}}0.609\\ (0.111)\end{tabular} &
  \begin{tabular}[c]{@{}c@{}}0.848\\ (0.108)\end{tabular} &
   &
  \begin{tabular}[c]{@{}c@{}}0.626\\ (0.111)\end{tabular} &
  \begin{tabular}[c]{@{}c@{}}0.691\\ (0.123)\end{tabular} &
  \begin{tabular}[c]{@{}c@{}}0.636\\ (0.116)\end{tabular} &
  \begin{tabular}[c]{@{}c@{}}0.755\\ (0.122)\end{tabular} \\
p=20 &
  \begin{tabular}[c]{@{}c@{}}0.987\\ (0.003)\end{tabular} &
  \begin{tabular}[c]{@{}c@{}}0.81\\ (0.072)\end{tabular} &
  \begin{tabular}[c]{@{}c@{}}0.671\\ (0.116)\end{tabular} &
  \begin{tabular}[c]{@{}c@{}}0.989\\ (0.003)\end{tabular} &
   &
  \begin{tabular}[c]{@{}c@{}}0.558\\ (0.065)\end{tabular} &
  \begin{tabular}[c]{@{}c@{}}0.545\\ (0.058)\end{tabular} &
  \begin{tabular}[c]{@{}c@{}}0.538\\ (0.05)\end{tabular} &
  \begin{tabular}[c]{@{}c@{}}0.685\\ (0.118)\end{tabular} &
   &
  \begin{tabular}[c]{@{}c@{}}0.532\\ (0.053)\end{tabular} &
  \begin{tabular}[c]{@{}c@{}}0.576\\ (0.084)\end{tabular} &
  \begin{tabular}[c]{@{}c@{}}0.542\\ (0.06)\end{tabular} &
  \begin{tabular}[c]{@{}c@{}}0.624\\ (0.103)\end{tabular} \\
p=50 &
  \begin{tabular}[c]{@{}c@{}}0.962\\ (0.006)\end{tabular} &
  \begin{tabular}[c]{@{}c@{}}0.426\\ (0.093)\end{tabular} &
  \begin{tabular}[c]{@{}c@{}}0.456\\ (0.047)\end{tabular} &
  \begin{tabular}[c]{@{}c@{}}0.971\\ (0.005)\end{tabular} &
   &
  \begin{tabular}[c]{@{}c@{}}0.512\\ (0.019)\end{tabular} &
  \begin{tabular}[c]{@{}c@{}}0.506\\ (0.021)\end{tabular} &
  \begin{tabular}[c]{@{}c@{}}0.505\\ (0.02)\end{tabular} &
  \begin{tabular}[c]{@{}c@{}}0.542\\ (0.046)\end{tabular} &
   &
  \begin{tabular}[c]{@{}c@{}}0.494\\ (0.018)\end{tabular} &
  \begin{tabular}[c]{@{}c@{}}0.508\\ (0.032)\end{tabular} &
  \begin{tabular}[c]{@{}c@{}}0.502\\ (0.022)\end{tabular} &
  \begin{tabular}[c]{@{}c@{}}0.511\\ (0.036)\end{tabular} \\ \hline
\end{tabular}
\end{table}

\subsubsection{Computational Time}\label{sec Computational Time}
We record the time consumption for implementing the distributed and global algorithms. To ensure impartiality, we utilize our manual R code when comparing the global SIR with the exact distributed estimation of SIR (E-DSIR). When comparing the global and approximate distributed algorithms, we employ the R package \emph{dr}. We record the running time of both the distributed and global algorithms under each model as the sample size grows. The data is generated in the standard normal distribution, where the dimension $p$ is taken to be $\{100, 200, 500\}$ for general moment-based inverse regression (SIR, SAVE and DR), and to be $\{500, 1000, 1500\}$ for comparison of the global SIR and E-DSIR. The number of worker nodes is set to be 5 and 10. The sample size grows from $1\times 10^4$ to $2\times10^5$. 

It is shown in Figure~\ref{fig time E-DSIR} and Figure~\ref{fig time SIR}, \ref{fig time SAVE}, \ref{fig time DR} in Appendix that as the sample size increases, the running time exhibits a roughly linear correlation with the growth in sample size. Besides, the larger the covariate dimension $p$, the longer the calculation time. More importantly, compared with the global algorithm, our distributed algorithm saves running time and reduces computational burden significantly. The efficiency further improves with an increase in number of nodes, resulting in shorter time consumption. When the sample size grows large enough, the implementation of the global algorithm is far more time-consuming, while our distributed algorithm still maintains a relatively modest computational cost. In general, the distributed algorithms substantially reduce the computational complexity. The efficiency of distributed computation is primarily influenced by the number of nodes, with little impact from the sample size.

\begin{figure}[H]
\centering
\includegraphics[scale=0.5]{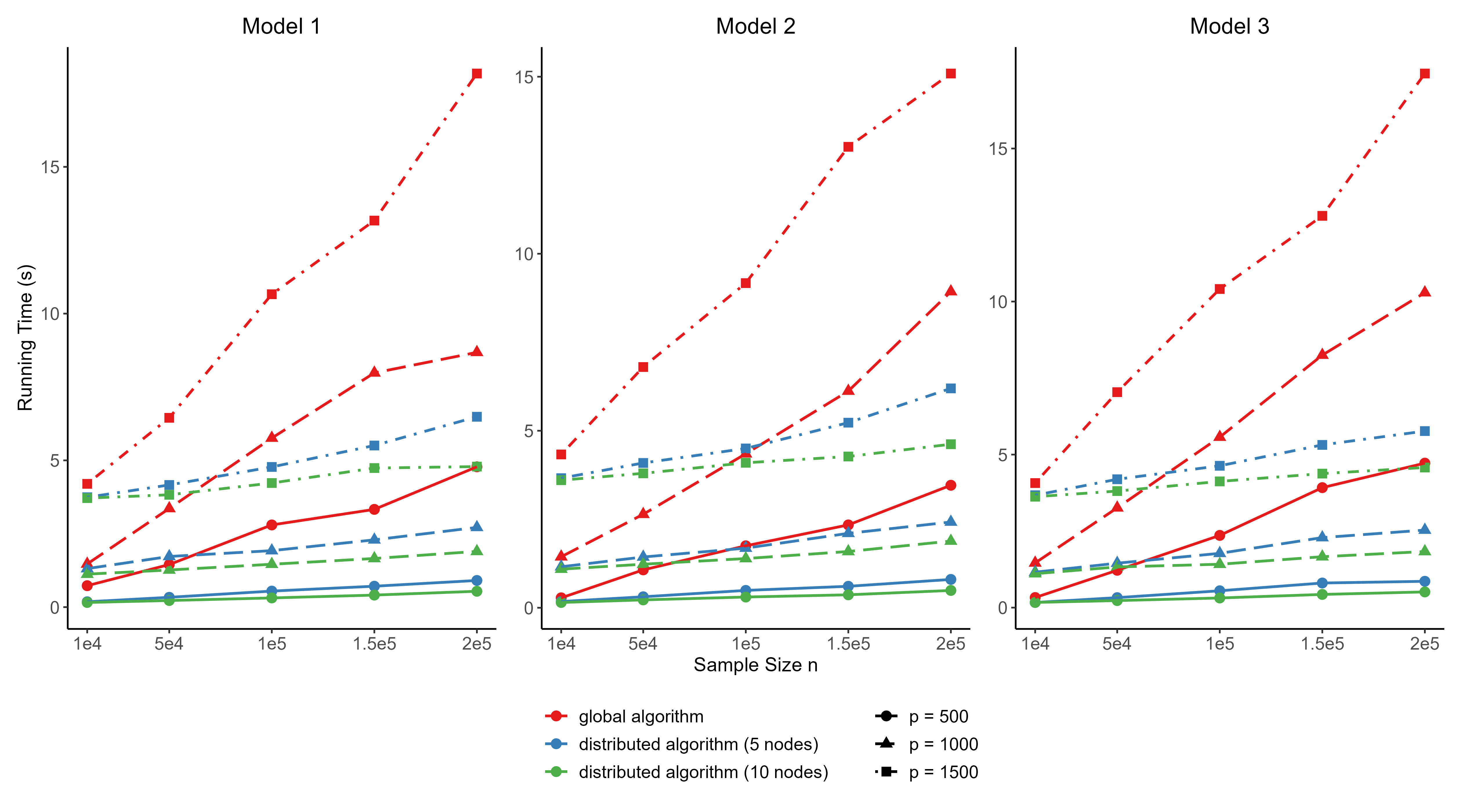}
\caption{\small Time consumption of E-DSIR and global SIR.}
\label{fig time E-DSIR}
\end{figure}

\subsubsection{Sensitivity Analysis}\label{sec sensitivity}

This section tests the effectiveness and robustness of the distributed algorithm under extreme conditions. Here we consider $n = 1000$ and $p = 100$. When observations are unevenly scattered across the distributed system, the sample size on some worker node is as few as 50, less than the predictor dimension. In this case, the SDR method locally fails. It casts a great challenge for the divide-and-conquer distributed algorithm.

We show the trace correlation and $R^2(\hat{\bm{\beta}})$ of the distributed and global SIR in Table~\ref{trace SIR p=100}, \ref{R2 SIR p=100} and Figure~\ref{fig trace SIR}. The distributed SIR performs comparable to the global SIR, and even outperforms the global SIR when data is heterogeneous on different nodes. The sensitivity analysis demonstrates that our proposed method is insensitive to extreme cases where the standard SDR method locally fails over some machines, and is robust enough and thus applicable to a wide range of distributed systems.

\begin{table}[H]
\setlength\tabcolsep{4pt}
\scriptsize
\caption{\small Trace correlation for distributed and global SIR algorithms when $p=100$.}
\label{trace SIR p=100}
\begin{tabular}{ccccccccccccccc}
\hline
 &
  \multicolumn{4}{c}{Standard Normal} &
   &
  \multicolumn{4}{c}{Heterogeneous Normal} &
   &
  \multicolumn{4}{c}{Dependent Normal} \\ \cline{2-5} \cline{7-10} \cline{12-15} 
 &
  \begin{tabular}[c]{@{}c@{}}Homo.\\ Equal.\end{tabular} &
  \begin{tabular}[c]{@{}c@{}}Hetero.\\ Equal.\end{tabular} &
  \begin{tabular}[c]{@{}c@{}}Hetero.\\ Unequal.\end{tabular} &
  \textbf{\emph{Global}} &
   &
  \begin{tabular}[c]{@{}c@{}}Homo.\\ Equal.\end{tabular} &
  \begin{tabular}[c]{@{}c@{}}Hetero.\\ Equal.\end{tabular} &
  \begin{tabular}[c]{@{}c@{}}Hetero.\\ Unequal.\end{tabular} &
  \textbf{\emph{Global}} &
   &
  \begin{tabular}[c]{@{}c@{}}Homo.\\ Equal.\end{tabular} &
  \begin{tabular}[c]{@{}c@{}}Hetero.\\ Equal.\end{tabular} &
  \begin{tabular}[c]{@{}c@{}}Hetero.\\ Unequal.\end{tabular} &
  \textbf{\emph{Global}}  \\ \hline
Model1 &
  \begin{tabular}[c]{@{}c@{}}0.998\\ (0)\end{tabular} &
  \begin{tabular}[c]{@{}c@{}}1\\ (0)\end{tabular} &
  \begin{tabular}[c]{@{}c@{}}1\\ (0)\end{tabular} &
  \begin{tabular}[c]{@{}c@{}}0.999\\ (0)\end{tabular} &
   &
  \begin{tabular}[c]{@{}c@{}}0.998\\ (0)\end{tabular} &
  \begin{tabular}[c]{@{}c@{}}1\\ (0)\end{tabular} &
  \begin{tabular}[c]{@{}c@{}}1\\ (0)\end{tabular} &
  \begin{tabular}[c]{@{}c@{}}0.999\\ (0)\end{tabular} &
   &
  \begin{tabular}[c]{@{}c@{}}0.99\\ (0.002)\end{tabular} &
  \begin{tabular}[c]{@{}c@{}}0.999\\ (0)\end{tabular} &
  \begin{tabular}[c]{@{}c@{}}0.999\\ (0)\end{tabular} &
  \begin{tabular}[c]{@{}c@{}}0.991\\ (0.002)\end{tabular} \\
Model2 &
  \begin{tabular}[c]{@{}c@{}}0.912\\ (0.012)\end{tabular} &
  \begin{tabular}[c]{@{}c@{}}0.956\\ (0.018)\end{tabular} &
  \begin{tabular}[c]{@{}c@{}}0.953\\ (0.019)\end{tabular} &
  \begin{tabular}[c]{@{}c@{}}0.93\\ (0.011)\end{tabular} &
   &
  \begin{tabular}[c]{@{}c@{}}0.504\\ (0.007)\end{tabular} &
  \begin{tabular}[c]{@{}c@{}}0.505\\ (0.007)\end{tabular} &
  \begin{tabular}[c]{@{}c@{}}0.505\\ (0.006)\end{tabular} &
  \begin{tabular}[c]{@{}c@{}}0.505\\ (0.008)\end{tabular} &
   &
  \begin{tabular}[c]{@{}c@{}}0.498\\ (0.002)\end{tabular} &
  \begin{tabular}[c]{@{}c@{}}0.501\\ (0.002)\end{tabular} &
  \begin{tabular}[c]{@{}c@{}}0.501\\ (0.002)\end{tabular} &
  \begin{tabular}[c]{@{}c@{}}0.499\\ (0.003)\end{tabular} \\
Model3 &
  \begin{tabular}[c]{@{}c@{}}0.967\\ (0.004)\end{tabular} &
  \begin{tabular}[c]{@{}c@{}}0.988\\ (0.003)\end{tabular} &
  \begin{tabular}[c]{@{}c@{}}0.982\\ (0.003)\end{tabular} &
  \begin{tabular}[c]{@{}c@{}}0.972\\ (0.004)\end{tabular} &
   &
  \begin{tabular}[c]{@{}c@{}}0.928\\ (0.011)\end{tabular} &
  \begin{tabular}[c]{@{}c@{}}0.987\\ (0.005)\end{tabular} &
  \begin{tabular}[c]{@{}c@{}}0.983\\ (0.006)\end{tabular} &
  \begin{tabular}[c]{@{}c@{}}0.943\\ (0.008)\end{tabular} &
   &
  \begin{tabular}[c]{@{}c@{}}0.845\\ (0.025)\end{tabular} &
  \begin{tabular}[c]{@{}c@{}}0.958\\ (0.017)\end{tabular} &
  \begin{tabular}[c]{@{}c@{}}0.947\\ (0.024)\end{tabular} &
  \begin{tabular}[c]{@{}c@{}}0.871\\ (0.017)\end{tabular} \\ \hline
\end{tabular}
\end{table}

\begin{table}[H]
\setlength\tabcolsep{4pt}
\scriptsize
\caption{\small $R^2(\hat{\bm{\beta}})$ for distributed and global SIR algorithms when $p=100$.}
\label{R2 SIR p=100}
\begin{tabular}{ccccccccccccccc}
\hline
 &
  \multicolumn{4}{c}{Standard Normal} &
   &
  \multicolumn{4}{c}{Heterogeneous Normal} &
   &
  \multicolumn{4}{c}{Dependent Normal} \\ \cline{2-5} \cline{7-10} \cline{12-15} 
 &
  \begin{tabular}[c]{@{}c@{}}Homo.\\ Equal.\end{tabular} &
  \begin{tabular}[c]{@{}c@{}}Hetero.\\ Equal.\end{tabular} &
  \begin{tabular}[c]{@{}c@{}}Hetero.\\ Unequal.\end{tabular} &
  \textbf{\emph{Global}} &
   &
  \begin{tabular}[c]{@{}c@{}}Homo.\\ Equal.\end{tabular} &
  \begin{tabular}[c]{@{}c@{}}Hetero.\\ Equal.\end{tabular} &
  \begin{tabular}[c]{@{}c@{}}Hetero.\\ Unequal.\end{tabular} &
  \textbf{\emph{Global}} &
   &
  \begin{tabular}[c]{@{}c@{}}Homo.\\ Equal.\end{tabular} &
  \begin{tabular}[c]{@{}c@{}}Hetero.\\ Equal.\end{tabular} &
  \begin{tabular}[c]{@{}c@{}}Hetero.\\ Unequal.\end{tabular} &
  \textbf{\emph{Global}}  \\ \hline
Model1 &
  \begin{tabular}[c]{@{}c@{}}0.998\\ (0)\end{tabular} &
  \begin{tabular}[c]{@{}c@{}}1\\ (0)\end{tabular} &
  \begin{tabular}[c]{@{}c@{}}1\\ (0)\end{tabular} &
  \begin{tabular}[c]{@{}c@{}}0.999\\ (0)\end{tabular} &
   &
  \begin{tabular}[c]{@{}c@{}}0.998\\ (0)\end{tabular} &
  \begin{tabular}[c]{@{}c@{}}1\\ (0)\end{tabular} &
  \begin{tabular}[c]{@{}c@{}}1\\ (0)\end{tabular} &
  \begin{tabular}[c]{@{}c@{}}0.999\\ (0)\end{tabular} &
   &
  \begin{tabular}[c]{@{}c@{}}0.998\\ (0)\end{tabular} &
  \begin{tabular}[c]{@{}c@{}}1\\ (0)\end{tabular} &
  \begin{tabular}[c]{@{}c@{}}1\\ (0)\end{tabular} &
  \begin{tabular}[c]{@{}c@{}}0.999\\ (0)\end{tabular} \\
Model2 &
  \begin{tabular}[c]{@{}c@{}}0.574\\ (0.017)\end{tabular} &
  \begin{tabular}[c]{@{}c@{}}0.524\\ (0.009)\end{tabular} &
  \begin{tabular}[c]{@{}c@{}}0.52\\ (0.014)\end{tabular} &
  \begin{tabular}[c]{@{}c@{}}0.585\\ (0.015)\end{tabular} &
   &
  \begin{tabular}[c]{@{}c@{}}0.998\\ (0)\end{tabular} &
  \begin{tabular}[c]{@{}c@{}}0.999\\ (0)\end{tabular} &
  \begin{tabular}[c]{@{}c@{}}0.999\\ (0)\end{tabular} &
  \begin{tabular}[c]{@{}c@{}}0.999\\ (0)\end{tabular} &
   &
  \begin{tabular}[c]{@{}c@{}}0.998\\ (0)\end{tabular} &
  \begin{tabular}[c]{@{}c@{}}1\\ (0)\end{tabular} &
  \begin{tabular}[c]{@{}c@{}}1\\ (0)\end{tabular} &
  \begin{tabular}[c]{@{}c@{}}0.999\\ (0)\end{tabular} \\
Model3 &
  \begin{tabular}[c]{@{}c@{}}0.988\\ (0.002)\end{tabular} &
  \begin{tabular}[c]{@{}c@{}}0.994\\ (0.002)\end{tabular} &
  \begin{tabular}[c]{@{}c@{}}0.991\\ (0.002)\end{tabular} &
  \begin{tabular}[c]{@{}c@{}}0.99\\ (0.002)\end{tabular} &
   &
  \begin{tabular}[c]{@{}c@{}}0.803\\ (0.012)\end{tabular} &
  \begin{tabular}[c]{@{}c@{}}0.782\\ (0.008)\end{tabular} &
  \begin{tabular}[c]{@{}c@{}}0.78\\ (0.009)\end{tabular} &
  \begin{tabular}[c]{@{}c@{}}0.798\\ (0.011)\end{tabular} &
   &
  \begin{tabular}[c]{@{}c@{}}0.788\\ (0.016)\end{tabular} &
  \begin{tabular}[c]{@{}c@{}}0.899\\ (0.012)\end{tabular} &
  \begin{tabular}[c]{@{}c@{}}0.887\\ (0.015)\end{tabular} &
  \begin{tabular}[c]{@{}c@{}}0.881\\ (0.016)\end{tabular} \\ \hline
\end{tabular}
\end{table}

\section{Real Data Analysis}\label{sec real}

We now apply the standard SIR method and our proposed distributed SIR algorithm to a real data set (\href{http://archive.ics.uci.edu/ml/datasets/Superconductivty+Data}{http://archive.ics.uci.edu/ml/datasets/Superconductivty+Data}). The data contains features of 21,263 superconductors. The response is the critical temperature, only at or below which do superconductors conduct current with zero resistance. 81 features are extracted from the chemical properties of the superconductor. Given the large dimension of predictors, SIR is a useful tool to reduce dimension without loss of information about the predictor, and meanwhile prevent overfitting. Before applying SIR, 80\% of the sample is randomly split as the training set, with the rest 20\% as the test set. Then we randomly scatter the observations to 5 workers in the distributed system. 

\begin{figure}[H]
\subfigure[Proportion of variance explained by global SIR]{ 
\begin{minipage}[t]{0.5\textwidth}
\centering
\includegraphics[scale=0.4]{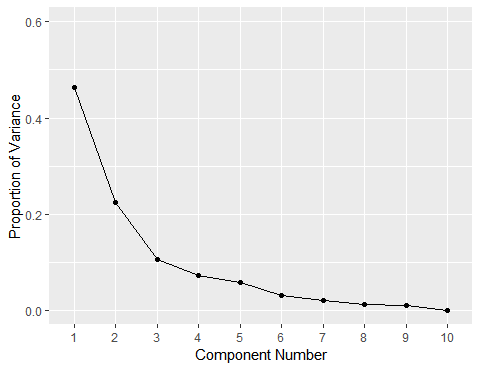}
\end{minipage}
}
\subfigure[Cumulative proportion explained by global SIR]{ 
\begin{minipage}[t]{0.5\textwidth}
\centering
\includegraphics[scale=0.4]{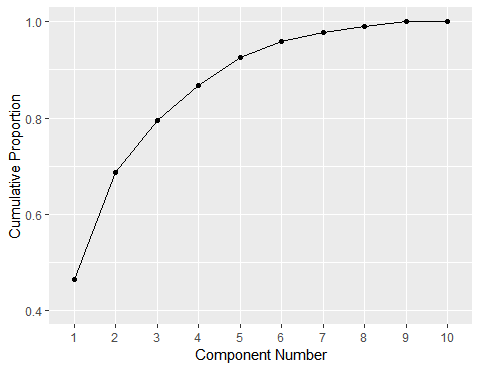}
\end{minipage}
}
\subfigure[Proportion of variance explained by distributed SIR]{ 
\begin{minipage}[t]{0.5\textwidth}
\centering
\includegraphics[scale=0.4]{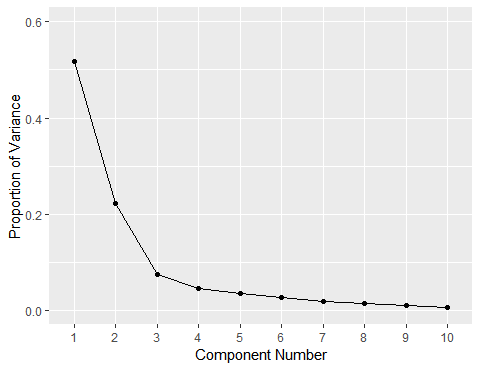}
\end{minipage}
}
\subfigure[Cumulative proportion explained by distributed SIR]{ 
\begin{minipage}[t]{0.5\textwidth}
\centering
\includegraphics[scale=0.4]{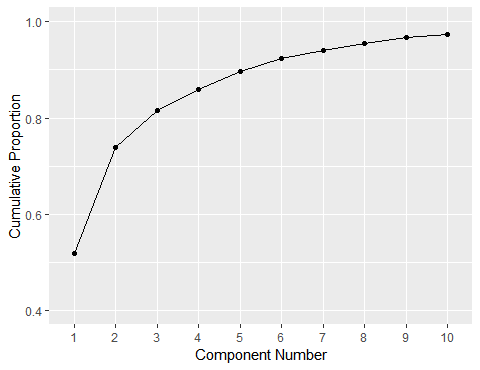}
\end{minipage}
}
\caption{\small Information explained by the first 10 directions of global and distributed SIR}
\label{pro and cum}
\end{figure}

The modeling process is to conduct sufficient dimension reduction in the first step, and then fit an ordinary least squares (OLS) regression model. In the training set, the original and distributed SIR algorithms are implemented respectively to estimate the SDR directions. Figure~\ref{pro and cum} presents the proportion and cumulative proportion of total variation explained by the top 10 directions from global SIR and distributed SIR. As the top 3 directions have already explained 79.5\% (in global SIR) and 81.5\% (in distributed SIR) of information contained in 81 covariates, we set the reduced dimension as $\hat{d} = 3$. The estimated SDR matrix (whose columns are SDR directions estimated from global or distributed SIR) is denoted as $\hat{\bm{B}}\in \mathbb{R}^{81\times3}$. The response $y_{train}$ in the training set and the reduced predictors $\hat{\bm{B}}^T \bm{X}_{train}$ are used to fit the OLS regression model. Detailed information of the model fitting is summarized in Table~\ref{lmfit.glb coef} and \ref{lmfit.dis coef}, which shows that the regression coefficients of the three reduced predictors obtained from global and distributed SIR are all significant. The regression diagnostics shown in Figure~\ref{lmfit.glb diag} and \ref{lmfit.dis diag} in the Appendix demonstrate that the fitted models meet the underlying statistical assumptions of OLS regression, i.e., linearity between predictors and the response, homoscedasticity, independence of observations, and normality of the distribution of the response for fixed predictors. 

\begin{table}[H]
\centering
\small
\caption{\small Regression coefficients of the reduced predictors from global SIR.}
\label{lmfit.glb coef}
\begin{tabular}{ccccc}
  \hline
 & Estimate & Std. Error & $t$-value & Pr($>|t|$) \\ 
  \hline
(Intercept) & 0.0033 & 0.0040 & 0.84 & 0.4000 \\ 
  Dir1 & -5.0014 & 0.0238 & -210.41 & \textless 2e-16 \\ 
  Dir2 & 4.9340 & 0.0943 & 52.30 & \textless 2e-16 \\ 
  Dir3 & -1.1039 & 0.0553 & -19.96 & \textless 2e-16 \\ 
   \hline
\end{tabular}
\end{table}

\begin{table}[H]
\centering
\small 
\caption{\small Regression coefficients of the reduced predictors from distributed SIR.}
\label{lmfit.dis coef}
\begin{tabular}{ccccc}
  \hline
 & Estimate & Std. Error & $t$-value & Pr($>|t|$) \\ 
  \hline
(Intercept) & 0.0036 & 0.0040 & 0.89 & 0.3710 \\ 
  Dir1 & -0.7833 & 0.1145 & -6.84 & 8.16e-12 \\ 
  Dir2 & 6.0586 & 0.0807 & 75.10 & \textless 2e-16 \\ 
  Dir3 & -1.7723 & 0.1301 & -13.62 & \textless 2e-16 \\ 
   \hline
\end{tabular}
\end{table}

Since the true SDR matrix $\bm{B}$ (whose columns span $\mathcal{S}_{Y|X}$) is unknown in the real data, we cannot continue to use the trace correlation or $R^2(\hat{\bm{\beta}})$ to evaluate the performance of SDR estimation. Instead, we apply $\hat{\bm{B}}$ and the OLS model obtained from the training set to the test set. The distance correlation \citep{szekely2007measuring} between $y_\text{test}$ and $\hat{\bm{B}}^T \bm{X}_\text{test}$, denoted as $\text{dcor}(y_\text{test}, \hat{\bm{B}}^T \bm{X}_\text{test})$, is employed to measure dependency between the response and the reduced predictors in the test set. The distance correlation ranges in $[0,1]$, and a larger value indicates the better performance. We also calculate the mean square error (MSE) of the prediction results in the test set, and a smaller value indicates the better performance. In addition, we report the goodness of fit $R^2$ and adjusted $R^2$ of the OLS model, which ranges in $[0,1]$ with the larger value indicating the better performance. We record the computational time of the global and distributed SIR as well.

The results of global and distributed SIR are concluded in Table~\ref{result real}. The reduced predictors obtained from the distributed SIR are closer to the response than the global SIR. Similar performance of the global and distributed SIR in model fitting and prediction implies that the distributed SIR achieves almost the same accuracy as the global SIR. In addition, the running time of distributed SIR is far less than global SIR, indicating that the distributed algorithm is much more efficient in computation when dealing with massive datasets.

\begin{table}[H]
\centering
\small
\caption{\small Results of global and distributed SIR applied to superconductor data.}
\label{result real}
\begin{tabular}{lccccc}
\hline
     & $\text{dcor}(y_\text{test}, \hat{\bm{B}}^T \bm{X}_\text{test})$ &  $R^2$  & Adjusted $R^2$  & MSE    & Running Time (s) \\ \cline{2-6} 
global SIR  & 0.8374         & 0.7360  & 0.7359        & 0.2613  & 1.660          \\
distributed SIR & 0.8467         & 0.7328 & 0.7327       & 0.2671   & 0.268       \\ \hline
\end{tabular}
\end{table}

\section{Conclusion and Discussion}\label{sec conclude}
This paper develops a unified framework for conditional-moment-based SDR, which involves two algorithms: exact distributed estimation of sliced inverse regression (E-DSIR) and approximate distributed estimation of general moment-based SDR. The first algorithm yields exactly the same estimation as the global SIR. Simple statistics are calculated and transmitted to reconstruct a global slicing on the master node, which requires two rounds of the map-reduce process. The second method extends the general moment-based inverse regression approaches to the distributed system, by conducting a one-shot approximation on the worker node and a weighted integration on the master node. The communication costs in these two approaches can be significantly reduced compared with transmitting the entire sample. 

The unified framework established in this paper allows samples to be heterogeneous and unbalanced on different nodes. It also explores rich details, including the selection of parameters, the relationship between local optimum and overall optimum under different circumstances, and the extreme case where the SDR method fails on some nodes. A wide range of simulation studies suggest that our proposed methods outperform the original SDR methods in computational efficiency while achieving similar estimation accuracy. Meanwhile, the distributed estimation is robust to different data generation schemes, different sample partition patterns, and some extreme cases. We illuminate the use of our proposed algorithms in a large-scale superconductor dataset concerning the prediction of critical temperature. Taking both the statistical accuracy and computational complexity into account, these two distributed algorithms are highly recommended for sufficient dimension reduction on massive datasets.

This research could be extended in many directions. For example, theoretical guarantees for the distributed estimator is under study. The global optimization of some parameters, such as the number of slices and the number of extracted eigenvectors on each server, can be further discussed. In addition, this framework is also expected to be applied to higher-order conditional moments, such as \citet{yin2002dimension}, \citet{yin2003estimating}, and so on.

\bibliographystyle{apalike}
\bibliography{reference}

\newpage
\section*{Appendix}

\begin{algorithm}[H] 
\caption{Original Algorithm of Sliced Inverse Regression} 
\label{alg:original SIR} 
\begin{algorithmic}[1] %1 means every line will show row number.
\REQUIRE ~~\\ %Input
Observations $\{(y_i, \bm{x}_i), i=1,...,n\}$;
\ENSURE ~~\\ %Output
\STATE Centralize $\bm{x}$ by computing $\bm{u}_i = \bm{x}_i - \bm{\bar{x}}$, where $\bm{\bar{x}}$ is the sample mean of $\bm{x}$.
\STATE Traverse $\{y_i, i=1,...,n\}$ to obtain the minimum response $y_\text{min}$ and the maximum response $y_\text{max}$.
\STATE Divide $\{y_i, i=1,...,n\}$ into $H$ slices $\{I_1,...,I_H\}$, where $I_h$ denotes the interval $(\tilde{y}_{h-1},\tilde{y}_h]$ for $h=1,...,H$, and $y_\text{min}=\tilde{y}_0<\tilde{y}_1<\cdots\tilde{y}_H=y_\text{max}$ are user-specified grid points.
\STATE Denote the sample size and the proportion of observations falling in slice $h$ respectively as $n_h$ and $\hat{p}_h = n_h/n$, $h = 1, ..., H$. Within each slice, calculate the sample mean of $\bm{u}_i$, denoted as $\hat{\bm{m}}_h=\sum_{y_i\in I_h}\bm{u}_i/n_h$.
\STATE Calculate the weighted covariance matrix $\hat{\bm{V}} = \sum_{h=1}^H \hat{p}_h \hat{\bm{m}}_h\hat{\bm{m}}_h^T$, and obtain the eigenvalues and eigenvectors for $\hat{\bm{V}}$. The corresponding eigenvectors of the top $K$ largest eigenvalues are denoted as $\hat{\bm{\eta}}_k$ ($k = 1, ..., K$).
\STATE \textbf{Return} estimates of SDR directions as $\hat{\bm{\beta}}_k=\hat{\bm{\Sigma}}^{-1}\hat{\bm{\eta}}_k$ ($k = 1, ..., K$), where $\hat{\bm{\Sigma}}=\frac{1}{n-1}\sum_{i=1}^n\bm{u}_i^T\bm{u}_i$ is the sample variance matrix.
\end{algorithmic}
\end{algorithm}

\newpage

\begin{table}[H]
\setlength\tabcolsep{4pt}
\scriptsize
\caption{\small Average (with standard deviation) of $R^2(\hat{\bm{\beta}})$ for distributed and global SAVE algorithms.}
\label{R2 SAVE}
\begin{tabular}{ccccccccccccccc}
\hline
 &
  \multicolumn{4}{c}{Standard Normal} &
   &
  \multicolumn{4}{c}{Heterogeneous Normal} &
   &
  \multicolumn{4}{c}{Dependent Normal} \\ \cline{2-5} \cline{7-10} \cline{12-15} 
 &
  \begin{tabular}[c]{@{}c@{}}Homo.\\ Equal.\end{tabular} &
  \begin{tabular}[c]{@{}c@{}}Hetero.\\ Equal.\end{tabular} &
  \begin{tabular}[c]{@{}c@{}}Hetero.\\ Unequal.\end{tabular} &
  \textbf{\emph{Global}} &
   &
  \begin{tabular}[c]{@{}c@{}}Homo.\\ Equal.\end{tabular} &
  \begin{tabular}[c]{@{}c@{}}Hetero.\\ Equal.\end{tabular} &
  \begin{tabular}[c]{@{}c@{}}Hetero.\\ Unequal.\end{tabular} &
  \textbf{\emph{Global}} &
   &
  \begin{tabular}[c]{@{}c@{}}Homo.\\ Equal.\end{tabular} &
  \begin{tabular}[c]{@{}c@{}}Hetero.\\ Equal.\end{tabular} &
  \begin{tabular}[c]{@{}c@{}}Hetero.\\ Unequal.\end{tabular} &
  \textbf{\emph{Global}}  \\ \hline
Model1 & & & & & & & & & & & & & & \\
p=10 &
  \begin{tabular}[c]{@{}c@{}}1\\ (0)\end{tabular} &
  \begin{tabular}[c]{@{}c@{}}1\\ (0)\end{tabular} &
  \begin{tabular}[c]{@{}c@{}}1\\ (0)\end{tabular} &
  \begin{tabular}[c]{@{}c@{}}1\\ (0)\end{tabular} &
   &
  \begin{tabular}[c]{@{}c@{}}1\\ (0)\end{tabular} &
  \begin{tabular}[c]{@{}c@{}}1\\ (0)\end{tabular} &
  \begin{tabular}[c]{@{}c@{}}1\\ (0)\end{tabular} &
  \begin{tabular}[c]{@{}c@{}}1\\ (0)\end{tabular} &
   &
  \begin{tabular}[c]{@{}c@{}}1\\ (0)\end{tabular} &
  \begin{tabular}[c]{@{}c@{}}1\\ (0)\end{tabular} &
  \begin{tabular}[c]{@{}c@{}}1\\ (0)\end{tabular} &
  \begin{tabular}[c]{@{}c@{}}1\\ (0)\end{tabular} \\
p=20 &
  \begin{tabular}[c]{@{}c@{}}1\\ (0)\end{tabular} &
  \begin{tabular}[c]{@{}c@{}}1\\ (0)\end{tabular} &
  \begin{tabular}[c]{@{}c@{}}0.998\\ (0.002)\end{tabular} &
  \begin{tabular}[c]{@{}c@{}}1\\ (0)\end{tabular} &
   &
  \begin{tabular}[c]{@{}c@{}}1\\ (0)\end{tabular} &
  \begin{tabular}[c]{@{}c@{}}1\\ (0)\end{tabular} &
  \begin{tabular}[c]{@{}c@{}}0.998\\ (0.002)\end{tabular} &
  \begin{tabular}[c]{@{}c@{}}1\\ (0)\end{tabular} &
   &
  \begin{tabular}[c]{@{}c@{}}1\\ (0)\end{tabular} &
  \begin{tabular}[c]{@{}c@{}}1\\ (0)\end{tabular} &
  \begin{tabular}[c]{@{}c@{}}1\\ (0)\end{tabular} &
  \begin{tabular}[c]{@{}c@{}}1\\ (0)\end{tabular} \\
p=50 &
  \begin{tabular}[c]{@{}c@{}}0.002\\ (0.003)\end{tabular} &
  \begin{tabular}[c]{@{}c@{}}0.06\\ (0.083)\end{tabular} &
  \begin{tabular}[c]{@{}c@{}}0.969\\ (0.013)\end{tabular} &
  \begin{tabular}[c]{@{}c@{}}0.999\\ (0)\end{tabular} &
   &
  \begin{tabular}[c]{@{}c@{}}0.002\\ (0.003)\end{tabular} &
  \begin{tabular}[c]{@{}c@{}}0.063\\ (0.085)\end{tabular} &
  \begin{tabular}[c]{@{}c@{}}0.968\\ (0.013)\end{tabular} &
  \begin{tabular}[c]{@{}c@{}}0.999\\ (0)\end{tabular} &
   &
  \begin{tabular}[c]{@{}c@{}}0.002\\ (0.003)\end{tabular} &
  \begin{tabular}[c]{@{}c@{}}0.036\\ (0.045)\end{tabular} &
  \begin{tabular}[c]{@{}c@{}}0.997\\ (0.002)\end{tabular} &
  \begin{tabular}[c]{@{}c@{}}0.999\\ (0)\end{tabular} \\ \hline
Model4  & & & & & & & & & & & & & & \\
p=10 &
  \begin{tabular}[c]{@{}c@{}}0.997\\ (0.001)\end{tabular} &
  \begin{tabular}[c]{@{}c@{}}0.971\\ (0.015)\end{tabular} &
  \begin{tabular}[c]{@{}c@{}}0.988\\ (0.005)\end{tabular} &
  \begin{tabular}[c]{@{}c@{}}0.997\\ (0.001)\end{tabular} &
   &
  \begin{tabular}[c]{@{}c@{}}1\\ (0)\end{tabular} &
  \begin{tabular}[c]{@{}c@{}}1\\ (0)\end{tabular} &
  \begin{tabular}[c]{@{}c@{}}1\\ (0)\end{tabular} &
  \begin{tabular}[c]{@{}c@{}}1\\ (0)\end{tabular} &
   &
  \begin{tabular}[c]{@{}c@{}}1\\ (0)\end{tabular} &
  \begin{tabular}[c]{@{}c@{}}1\\ (0)\end{tabular} &
  \begin{tabular}[c]{@{}c@{}}1\\ (0)\end{tabular} &
  \begin{tabular}[c]{@{}c@{}}1\\ (0)\end{tabular} \\
p=20 &
  \begin{tabular}[c]{@{}c@{}}0.993\\ (0.002)\end{tabular} &
  \begin{tabular}[c]{@{}c@{}}0.914\\ (0.04)\end{tabular} &
  \begin{tabular}[c]{@{}c@{}}0.974\\ (0.011)\end{tabular} &
  \begin{tabular}[c]{@{}c@{}}0.994\\ (0.002)\end{tabular} &
   &
  \begin{tabular}[c]{@{}c@{}}1\\ (0)\end{tabular} &
  \begin{tabular}[c]{@{}c@{}}1\\ (0)\end{tabular} &
  \begin{tabular}[c]{@{}c@{}}0.997\\ (0.002)\end{tabular} &
  \begin{tabular}[c]{@{}c@{}}1\\ (0)\end{tabular} &
   &
  \begin{tabular}[c]{@{}c@{}}0.999\\ (0)\end{tabular} &
  \begin{tabular}[c]{@{}c@{}}0.999\\ (0)\end{tabular} &
  \begin{tabular}[c]{@{}c@{}}0.999\\ (0.001)\end{tabular} &
  \begin{tabular}[c]{@{}c@{}}1\\ (0)\end{tabular} \\
p=50 &
  \begin{tabular}[c]{@{}c@{}}0.975\\ (0.005)\end{tabular} &
  \begin{tabular}[c]{@{}c@{}}0.724\\ (0.144)\end{tabular} &
  \begin{tabular}[c]{@{}c@{}}0.912\\ (0.032)\end{tabular} &
  \begin{tabular}[c]{@{}c@{}}0.982\\ (0.004)\end{tabular} &
   &
  \begin{tabular}[c]{@{}c@{}}0.002\\ (0.003)\end{tabular} &
  \begin{tabular}[c]{@{}c@{}}0.497\\ (0.379)\end{tabular} &
  \begin{tabular}[c]{@{}c@{}}0.906\\ (0.059)\end{tabular} &
  \begin{tabular}[c]{@{}c@{}}0.999\\ (0)\end{tabular} &
   &
  \begin{tabular}[c]{@{}c@{}}0.002\\ (0.003)\end{tabular} &
  \begin{tabular}[c]{@{}c@{}}0.796\\ (0.311)\end{tabular} &
  \begin{tabular}[c]{@{}c@{}}0.975\\ (0.026)\end{tabular} &
  \begin{tabular}[c]{@{}c@{}}0.999\\ (0)\end{tabular} \\ \hline
Model5 & & & & & & & & & & & & & & \\
p=10 &
  \begin{tabular}[c]{@{}c@{}}0.996\\ (0.002)\end{tabular} &
  \begin{tabular}[c]{@{}c@{}}0.977\\ (0.013)\end{tabular} &
  \begin{tabular}[c]{@{}c@{}}0.985\\ (0.01)\end{tabular} &
  \begin{tabular}[c]{@{}c@{}}0.998\\ (0.001)\end{tabular} &
   &
  \begin{tabular}[c]{@{}c@{}}1\\ (0)\end{tabular} &
  \begin{tabular}[c]{@{}c@{}}1\\ (0)\end{tabular} &
  \begin{tabular}[c]{@{}c@{}}1\\ (0)\end{tabular} &
  \begin{tabular}[c]{@{}c@{}}1\\ (0)\end{tabular} &
   &
  \begin{tabular}[c]{@{}c@{}}1\\ (0)\end{tabular} &
  \begin{tabular}[c]{@{}c@{}}1\\ (0)\end{tabular} &
  \begin{tabular}[c]{@{}c@{}}1\\ (0)\end{tabular} &
  \begin{tabular}[c]{@{}c@{}}1\\ (0)\end{tabular} \\
p=20 &
  \begin{tabular}[c]{@{}c@{}}0.954\\ (0.022)\end{tabular} &
  \begin{tabular}[c]{@{}c@{}}0.92\\ (0.038)\end{tabular} &
  \begin{tabular}[c]{@{}c@{}}0.959\\ (0.022)\end{tabular} &
  \begin{tabular}[c]{@{}c@{}}0.996\\ (0.001)\end{tabular} &
   &
  \begin{tabular}[c]{@{}c@{}}1\\ (0)\end{tabular} &
  \begin{tabular}[c]{@{}c@{}}1\\ (0)\end{tabular} &
  \begin{tabular}[c]{@{}c@{}}0.998\\ (0.002)\end{tabular} &
  \begin{tabular}[c]{@{}c@{}}1\\ (0)\end{tabular} &
   &
  \begin{tabular}[c]{@{}c@{}}1\\ (0)\end{tabular} &
  \begin{tabular}[c]{@{}c@{}}1\\ (0)\end{tabular} &
  \begin{tabular}[c]{@{}c@{}}1\\ (0)\end{tabular} &
  \begin{tabular}[c]{@{}c@{}}1\\ (0)\end{tabular} \\
p=50 &
  \begin{tabular}[c]{@{}c@{}}0.021\\ (0.043)\end{tabular} &
  \begin{tabular}[c]{@{}c@{}}0.655\\ (0.159)\end{tabular} &
  \begin{tabular}[c]{@{}c@{}}0.88\\ (0.084)\end{tabular} &
  \begin{tabular}[c]{@{}c@{}}0.977\\ (0.006)\end{tabular} &
   &
  \begin{tabular}[c]{@{}c@{}}0.003\\ (0.004)\end{tabular} &
  \begin{tabular}[c]{@{}c@{}}0.054\\ (0.088)\end{tabular} &
  \begin{tabular}[c]{@{}c@{}}0.968\\ (0.014)\end{tabular} &
  \begin{tabular}[c]{@{}c@{}}0.999\\ (0)\end{tabular} &
   &
  \begin{tabular}[c]{@{}c@{}}0.003\\ (0.004)\end{tabular} &
  \begin{tabular}[c]{@{}c@{}}0.074\\ (0.181)\end{tabular} &
  \begin{tabular}[c]{@{}c@{}}0.992\\ (0.005)\end{tabular} &
  \begin{tabular}[c]{@{}c@{}}0.999\\ (0)\end{tabular} \\ \hline
\end{tabular}
\end{table}

\begin{table}[H]
\setlength\tabcolsep{4pt}
\scriptsize
\caption{\small Average (with standard deviation) of $R^2(\hat{\bm{\beta}})$ for distributed and global DR algorithms.}
\label{R2 DR}
\begin{tabular}{ccccccccccccccc}
\hline
 &
  \multicolumn{4}{c}{Standard Normal} &
   &
  \multicolumn{4}{c}{Heterogeneous Normal} &
   &
  \multicolumn{4}{c}{Dependent Normal} \\ \cline{2-5} \cline{7-10} \cline{12-15} 
 &
  \begin{tabular}[c]{@{}c@{}}Homo.\\ Equal.\end{tabular} &
  \begin{tabular}[c]{@{}c@{}}Hetero.\\ Equal.\end{tabular} &
  \begin{tabular}[c]{@{}c@{}}Hetero.\\ Unequal.\end{tabular} &
  \textbf{\emph{Global}} &
   &
  \begin{tabular}[c]{@{}c@{}}Homo.\\ Equal.\end{tabular} &
  \begin{tabular}[c]{@{}c@{}}Hetero.\\ Equal.\end{tabular} &
  \begin{tabular}[c]{@{}c@{}}Hetero.\\ Unequal.\end{tabular} &
  \textbf{\emph{Global}} &
   &
  \begin{tabular}[c]{@{}c@{}}Homo.\\ Equal.\end{tabular} &
  \begin{tabular}[c]{@{}c@{}}Hetero.\\ Equal.\end{tabular} &
  \begin{tabular}[c]{@{}c@{}}Hetero.\\ Unequal.\end{tabular} &
  \textbf{\emph{Global}}  \\ \hline
Model6 & & & & & & & & & & & & & & \\
p=10 &
  \begin{tabular}[c]{@{}c@{}}0.986\\ (0.006)\end{tabular} &
  \begin{tabular}[c]{@{}c@{}}0.869\\ (0.096)\end{tabular} &
  \begin{tabular}[c]{@{}c@{}}0.78\\ (0.126)\end{tabular} &
  \begin{tabular}[c]{@{}c@{}}0.987\\ (0.006)\end{tabular} &
   &
  \begin{tabular}[c]{@{}c@{}}0.993\\ (0.002)\end{tabular} &
  \begin{tabular}[c]{@{}c@{}}0.992\\ (0.003)\end{tabular} &
  \begin{tabular}[c]{@{}c@{}}0.992\\ (0.003)\end{tabular} &
  \begin{tabular}[c]{@{}c@{}}0.993\\ (0.002)\end{tabular} &
   &
  \begin{tabular}[c]{@{}c@{}}0.964\\ (0.004)\end{tabular} &
  \begin{tabular}[c]{@{}c@{}}0.981\\ (0.004)\end{tabular} &
  \begin{tabular}[c]{@{}c@{}}0.982\\ (0.004)\end{tabular} &
  \begin{tabular}[c]{@{}c@{}}0.961\\ (0.004)\end{tabular} \\
p=20 &
  \begin{tabular}[c]{@{}c@{}}0.981\\ (0.007)\end{tabular} &
  \begin{tabular}[c]{@{}c@{}}0.679\\ (0.155)\end{tabular} &
  \begin{tabular}[c]{@{}c@{}}0.673\\ (0.124)\end{tabular} &
  \begin{tabular}[c]{@{}c@{}}0.981\\ (0.006)\end{tabular} &
   &
  \begin{tabular}[c]{@{}c@{}}0.992\\ (0.002)\end{tabular} &
  \begin{tabular}[c]{@{}c@{}}0.989\\ (0.003)\end{tabular} &
  \begin{tabular}[c]{@{}c@{}}0.989\\ (0.003)\end{tabular} &
  \begin{tabular}[c]{@{}c@{}}0.992\\ (0.002)\end{tabular} &
   &
  \begin{tabular}[c]{@{}c@{}}0.96\\ (0.004)\end{tabular} &
  \begin{tabular}[c]{@{}c@{}}0.98\\ (0.005)\end{tabular} &
  \begin{tabular}[c]{@{}c@{}}0.982\\ (0.004)\end{tabular} &
  \begin{tabular}[c]{@{}c@{}}0.964\\ (0.004)\end{tabular} \\
p=50 &
  \begin{tabular}[c]{@{}c@{}}0.959\\ (0.01)\end{tabular} &
  \begin{tabular}[c]{@{}c@{}}0.212\\ (0.154)\end{tabular} &
  \begin{tabular}[c]{@{}c@{}}0.309\\ (0.142)\end{tabular} &
  \begin{tabular}[c]{@{}c@{}}0.968\\ (0.006)\end{tabular} &
   &
  \begin{tabular}[c]{@{}c@{}}0.988\\ (0.002)\end{tabular} &
  \begin{tabular}[c]{@{}c@{}}0.979\\ (0.004)\end{tabular} &
  \begin{tabular}[c]{@{}c@{}}0.979\\ (0.004)\end{tabular} &
  \begin{tabular}[c]{@{}c@{}}0.988\\ (0.002)\end{tabular} &
   &
  \begin{tabular}[c]{@{}c@{}}0.962\\ (0.004)\end{tabular} &
  \begin{tabular}[c]{@{}c@{}}0.978\\ (0.005)\end{tabular} &
  \begin{tabular}[c]{@{}c@{}}0.976\\ (0.005)\end{tabular} &
  \begin{tabular}[c]{@{}c@{}}0.962\\ (0.004)\end{tabular} \\ \hline
Model7 & & & & & & & & & & & & & & \\
p=10 &
  \begin{tabular}[c]{@{}c@{}}0.99\\ (0.035)\end{tabular} &
  \begin{tabular}[c]{@{}c@{}}0.797\\ (0.156)\end{tabular} &
  \begin{tabular}[c]{@{}c@{}}0.885\\ (0.112)\end{tabular} &
  \begin{tabular}[c]{@{}c@{}}0.99\\ (0.038)\end{tabular} &
   &
  \begin{tabular}[c]{@{}c@{}}0.99\\ (0.05)\end{tabular} &
  \begin{tabular}[c]{@{}c@{}}0.989\\ (0.043)\end{tabular} &
  \begin{tabular}[c]{@{}c@{}}0.986\\ (0.06)\end{tabular} &
  \begin{tabular}[c]{@{}c@{}}0.989\\ (0.042)\end{tabular} &
   &
  \begin{tabular}[c]{@{}c@{}}0.973\\ (0.044)\end{tabular} &
  \begin{tabular}[c]{@{}c@{}}0.986\\ (0.043)\end{tabular} &
  \begin{tabular}[c]{@{}c@{}}0.985\\ (0.036)\end{tabular} &
  \begin{tabular}[c]{@{}c@{}}0.97\\ (0.042)\end{tabular} \\
p=20 &
  \begin{tabular}[c]{@{}c@{}}0.986\\ (0.037)\end{tabular} &
  \begin{tabular}[c]{@{}c@{}}0.698\\ (0.168)\end{tabular} &
  \begin{tabular}[c]{@{}c@{}}0.784\\ (0.149)\end{tabular} &
  \begin{tabular}[c]{@{}c@{}}0.986\\ (0.031)\end{tabular} &
   &
  \begin{tabular}[c]{@{}c@{}}0.988\\ (0.044)\end{tabular} &
  \begin{tabular}[c]{@{}c@{}}0.992\\ (0.011)\end{tabular} &
  \begin{tabular}[c]{@{}c@{}}0.989\\ (0.025)\end{tabular} &
  \begin{tabular}[c]{@{}c@{}}0.99\\ (0.039)\end{tabular} &
   &
  \begin{tabular}[c]{@{}c@{}}0.975\\ (0.019)\end{tabular} &
  \begin{tabular}[c]{@{}c@{}}0.987\\ (0.028)\end{tabular} &
  \begin{tabular}[c]{@{}c@{}}0.989\\ (0.017)\end{tabular} &
  \begin{tabular}[c]{@{}c@{}}0.974\\ (0.028)\end{tabular} \\
p=50 &
  \begin{tabular}[c]{@{}c@{}}0.967\\ (0.043)\end{tabular} &
  \begin{tabular}[c]{@{}c@{}}0.43\\ (0.227)\end{tabular} &
  \begin{tabular}[c]{@{}c@{}}0.512\\ (0.193)\end{tabular} &
  \begin{tabular}[c]{@{}c@{}}0.974\\ (0.028)\end{tabular} &
   &
  \begin{tabular}[c]{@{}c@{}}0.982\\ (0.044)\end{tabular} &
  \begin{tabular}[c]{@{}c@{}}0.98\\ (0.029)\end{tabular} &
  \begin{tabular}[c]{@{}c@{}}0.98\\ (0.03)\end{tabular} &
  \begin{tabular}[c]{@{}c@{}}0.987\\ (0.038)\end{tabular} &
   &
  \begin{tabular}[c]{@{}c@{}}0.97\\ (0.031)\end{tabular} &
  \begin{tabular}[c]{@{}c@{}}0.982\\ (0.042)\end{tabular} &
  \begin{tabular}[c]{@{}c@{}}0.984\\ (0.026)\end{tabular} &
  \begin{tabular}[c]{@{}c@{}}0.97\\ (0.043)\end{tabular} \\ \hline
Model8 & & & & & & & & & & & & & & \\
p=10 &
  \begin{tabular}[c]{@{}c@{}}0.97\\ (0.025)\end{tabular} &
  \begin{tabular}[c]{@{}c@{}}0.932\\ (0.06)\end{tabular} &
  \begin{tabular}[c]{@{}c@{}}0.96\\ (0.024)\end{tabular} &
  \begin{tabular}[c]{@{}c@{}}0.947\\ (0.051)\end{tabular} &
   &
  \begin{tabular}[c]{@{}c@{}}0.98\\ (0.003)\end{tabular} &
  \begin{tabular}[c]{@{}c@{}}0.972\\ (0.007)\end{tabular} &
  \begin{tabular}[c]{@{}c@{}}0.978\\ (0.005)\end{tabular} &
  \begin{tabular}[c]{@{}c@{}}0.979\\ (0.004)\end{tabular} &
   &
  \begin{tabular}[c]{@{}c@{}}0.989\\ (0.001)\end{tabular} &
  \begin{tabular}[c]{@{}c@{}}0.992\\ (0.002)\end{tabular} &
  \begin{tabular}[c]{@{}c@{}}0.992\\ (0.002)\end{tabular} &
  \begin{tabular}[c]{@{}c@{}}0.989\\ (0.002)\end{tabular} \\
p=20 &
  \begin{tabular}[c]{@{}c@{}}0.976\\ (0.018)\end{tabular} &
  \begin{tabular}[c]{@{}c@{}}0.864\\ (0.083)\end{tabular} &
  \begin{tabular}[c]{@{}c@{}}0.92\\ (0.035)\end{tabular} &
  \begin{tabular}[c]{@{}c@{}}0.958\\ (0.032)\end{tabular} &
   &
  \begin{tabular}[c]{@{}c@{}}0.978\\ (0.004)\end{tabular} &
  \begin{tabular}[c]{@{}c@{}}0.971\\ (0.006)\end{tabular} &
  \begin{tabular}[c]{@{}c@{}}0.972\\ (0.006)\end{tabular} &
  \begin{tabular}[c]{@{}c@{}}0.977\\ (0.004)\end{tabular} &
   &
  \begin{tabular}[c]{@{}c@{}}0.989\\ (0.002)\end{tabular} &
  \begin{tabular}[c]{@{}c@{}}0.99\\ (0.003)\end{tabular} &
  \begin{tabular}[c]{@{}c@{}}0.991\\ (0.002)\end{tabular} &
  \begin{tabular}[c]{@{}c@{}}0.989\\ (0.001)\end{tabular} \\
p=50 &
  \begin{tabular}[c]{@{}c@{}}0.977\\ (0.007)\end{tabular} &
  \begin{tabular}[c]{@{}c@{}}0.604\\ (0.152)\end{tabular} &
  \begin{tabular}[c]{@{}c@{}}0.813\\ (0.056)\end{tabular} &
  \begin{tabular}[c]{@{}c@{}}0.955\\ (0.026)\end{tabular} &
   &
  \begin{tabular}[c]{@{}c@{}}0.973\\ (0.004)\end{tabular} &
  \begin{tabular}[c]{@{}c@{}}0.96\\ (0.008)\end{tabular} &
  \begin{tabular}[c]{@{}c@{}}0.961\\ (0.007)\end{tabular} &
  \begin{tabular}[c]{@{}c@{}}0.974\\ (0.004)\end{tabular} &
   &
  \begin{tabular}[c]{@{}c@{}}0.987\\ (0.002)\end{tabular} &
  \begin{tabular}[c]{@{}c@{}}0.986\\ (0.003)\end{tabular} &
  \begin{tabular}[c]{@{}c@{}}0.986\\ (0.003)\end{tabular} &
  \begin{tabular}[c]{@{}c@{}}0.987\\ (0.002)\end{tabular} \\ \hline
\end{tabular}
\end{table}

\begin{figure}[H]
\centering
\includegraphics[scale=0.6]{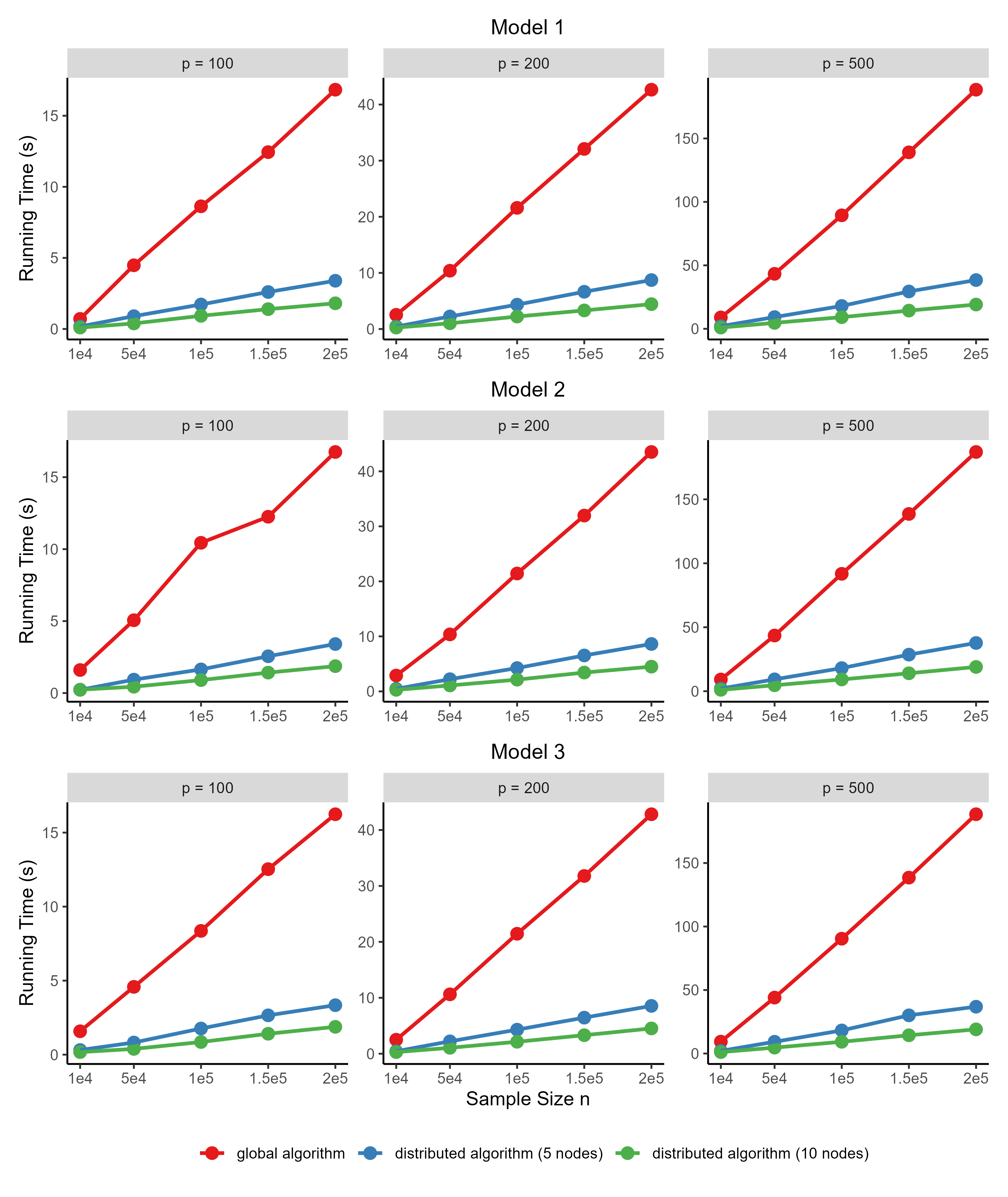}
\caption{\small Time consumption of distributed and global SIR.}
\label{fig time SIR}
\end{figure}

\begin{figure}[H]
\centering
\includegraphics[scale=0.6]{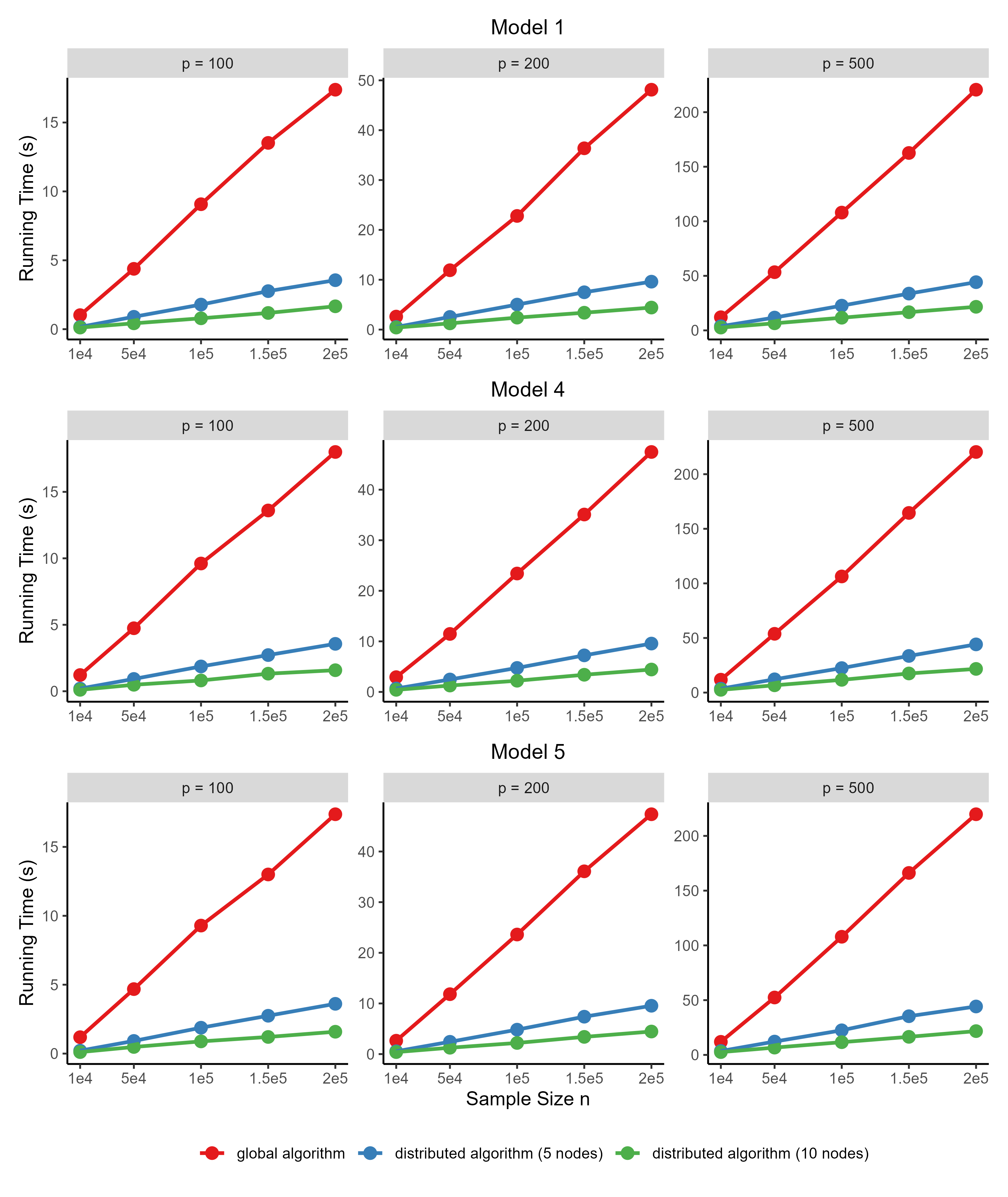}
\caption{\small Time consumption of distributed and global SAVE.}
\label{fig time SAVE}
\end{figure}

\begin{figure}[H]
\centering
\includegraphics[scale=0.6]{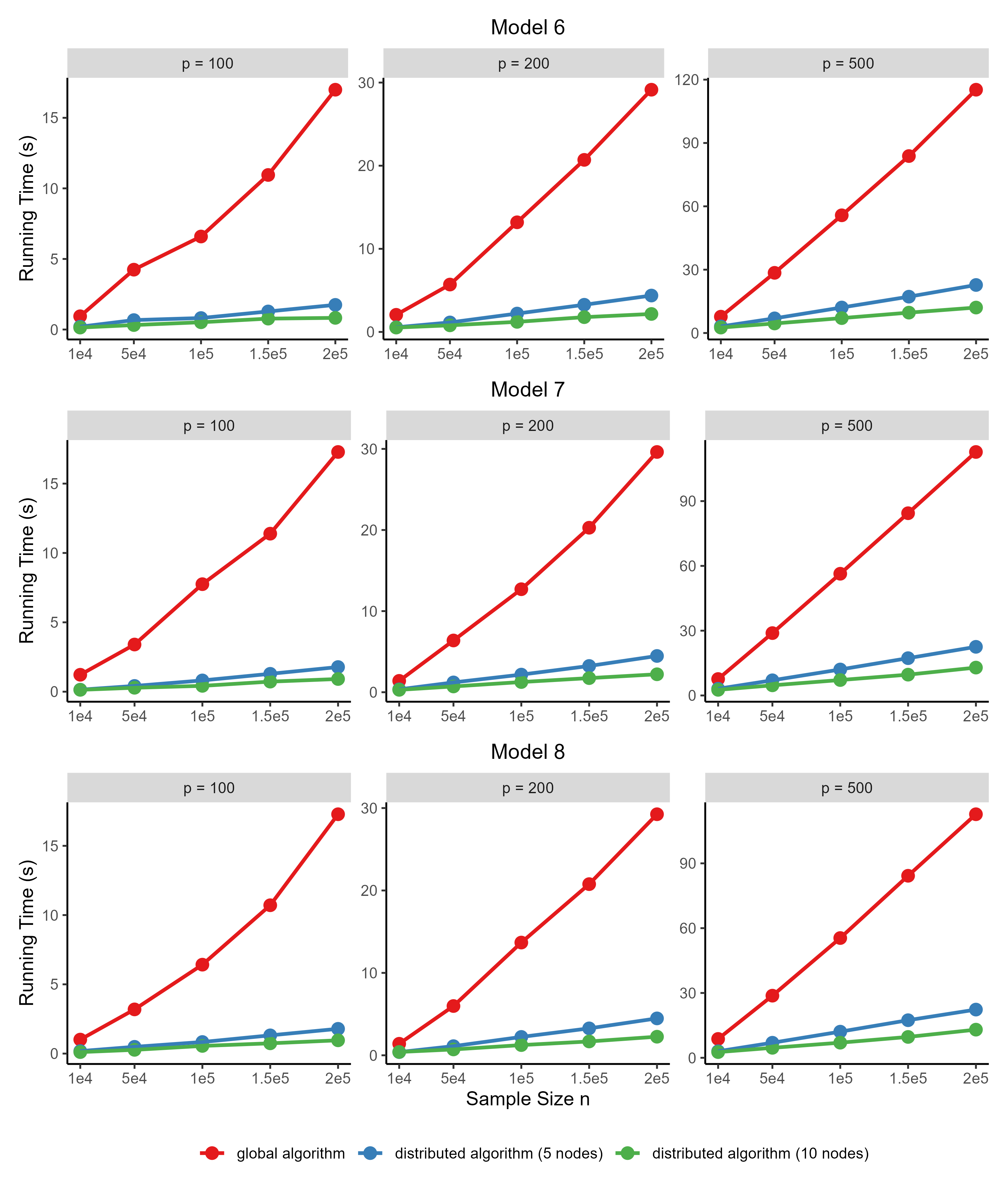}
\caption{\small Time consumption of distributed and global DR.}
\label{fig time DR}
\end{figure}

\begin{figure}[H]
\centering
\includegraphics[scale=0.8]{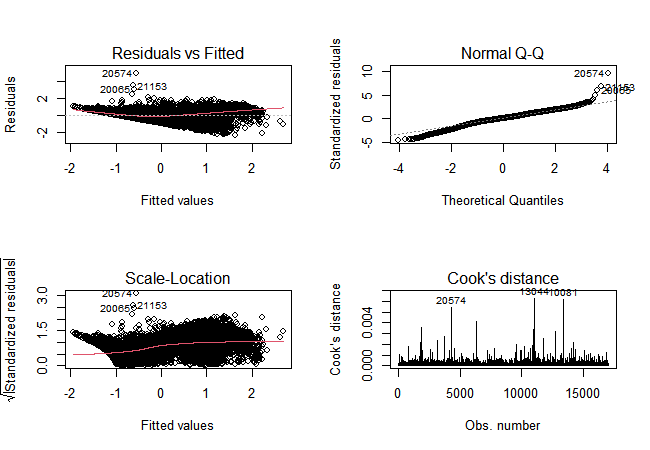}
\caption{\small Regression diagnostics of OLS model based on global SIR.}
\label{lmfit.glb diag}
\end{figure}

\begin{figure}[H]
\centering
\includegraphics[scale=0.8]{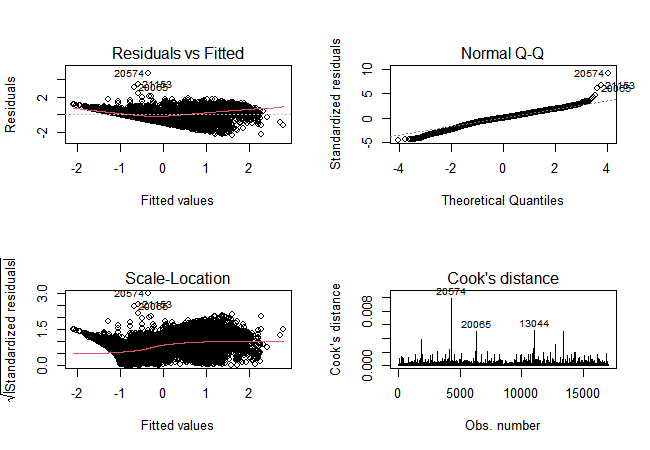}
\caption{\small Regression diagnostics of OLS model based on distributed SIR.}
\label{lmfit.dis diag}
\end{figure}

\end{document}